\documentclass[notitlepage,aps,preprintnumbers,onecolumn,amsmath,amssymb,floatfix,pra]{revtex4-1}

\pdfoutput=1
\usepackage[utf8]{inputenc}

\usepackage{amsmath}
\usepackage{amsfonts}
\usepackage{braket}
\usepackage{comment}
\usepackage{graphicx}
\usepackage[usenames,dvipsnames,svgnames,table]{xcolor}

\usepackage[colorlinks=true,linkcolor=blue, citecolor=blue, bookmarks]{hyperref}
\usepackage{tikz,pgfplots}
\usepackage{wasysym}

\newcommand{\rhop}{\rho}

\newcommand{\Sred}{S_{\mathrm{red}}}
\newcommand{\SYY}{S_{\mathrm{YY}}}

\newcommand{\Srat}{R} 	

\newcommand{\im}{\mathrm{i}}

\newcommand{\nn}{\nonumber \\}
\newcommand{\ignore}[1]{{}}
\newcommand{\be}{\begin{equation}}
\newcommand{\ee}{\end{equation}}
\def\doi{http://dx.doi.org/}

\tikzset{
	yyentropy-data-plot/.pic = {
		\begin{axis}[width=11cm,height=7.6cm,
			xmin=-3.14159265359,xmax=3.14159265359,
			ymin=0,ymax=0.125
			]
			\addplot[domain=-3.14159265359:3.14159265359,red,thick] table {\yyet};
			\addplot[domain=-3.14159265359:3.14159265359,dodgerblue,thick] table {\yyef};
			\addplot[domain=-3.14159265359:3.14159265359,lightseagreen,thick] table {\yyes};
		\end{axis}
	}
}

\tikzset{
    yyentropy-legend/.pic = {
		\draw[red, thick] (0,0.7) -- (1,0.7);
		\draw  node[right] at (-1,0.7) {$\nu=2$};
		\draw[dodgerblue, thick] (0,0.35) -- (1,0.35);
		\draw  node[right] at (-1,0.35) {$\nu=5$};
		\draw[lightseagreen, thick] (0,0) -- (1,0);
		\draw  node[right] at (-1,0) {$\nu=6$};
    }
}

\definecolor{lightseagreen}{RGB}{60,179,113}
\definecolor{dodgerblue}{RGB}{30,144,255}

\begin{document}
\title{Entanglement and diagonal entropies after a quench with no pair structure}%
\author{Bruno Bertini}
\address{Department of Physics, Faculty of Mathematics and Physics,
University of Ljubljana, Jadranska 19, SI-1000 Ljubljana, Slovenia}
\author{Elena Tartaglia}
\address{SISSA and INFN, via Bonomea 265, 34136, Trieste, Italy}
\author{Pasquale Calabrese}
\address{SISSA and INFN, via Bonomea 265, 34136, Trieste, Italy}
\address{International Centre for Theoretical Physics (ICTP), I-34151, Trieste, Italy}

\begin{abstract}
A typical working condition in the study of quantum quenches is that the initial state produces a distribution of quasiparticle excitations with an opposite-momentum-pair structure. In this work we investigate the dynamical and stationary properties of the entanglement entropy after a quench from initial states which do not have such structure: instead of pairs of excitations they generate $\nu$-plets of correlated excitations with $\nu>2$. Our study is carried out focusing on a system of non-interacting fermions on the lattice. We study the time evolution of the entanglement entropy showing that the standard semiclassical formula is not applicable. We propose a suitable generalisation which correctly describes the entanglement entropy evolution and perfectly matches numerical data. We finally consider the relation between the thermodynamic entropy of the stationary state and the diagonal entropy, showing that when there is no pair structure their ratio depends on the details of the initial state and lies generically between $1/2$ and $1$. 
\end{abstract}
\maketitle

\section{Introduction}

Understanding how entanglement and correlations spread in out-of-equilibrium many-body quantum systems is 
an extremely fascinating 
topic which turned out to be of fundamental importance for several 
problems in condensed matter, statistical physics and quantum field theory. 
In this manuscript we 
focus on global quenches,
in which an isolated many-body quantum system is prepared at $t=0$ in a finite-energy-density pure state $|\psi_0\rangle$, 
and for $t>0$ it is evolved unitarily with dynamics governed by a Hamiltonian $H$ (see the reviews~\cite{pssv-11,gogolin-2015,vidmar-2016,calabrese-2016,cc-16,ef-16,quasilocal-16}). 
The theoretical study of quench problems has been boosted by seminal cold-atom experiments~\cite{langen-rev,kinoshita-2006,hofferberth-2007,
trotzky-2012,gring-2012,cheneau-2012,langen-2013,fukuhara-2013,langen-2015,
bsjs-18} probing many different aspects of the non-equilibrium unitary dynamics of quantum systems.

If a many-body quantum system is in a pure state $|\psi\rangle$, the bipartite entanglement between a subsystem $A$ and its complement 
$\bar A$ may be measured by the entanglement entropy \cite{faziorev,CCrev,Eisrev,Lafrev}.
This is defined as the von Neumann entropy of the reduced density matrix $\boldsymbol \rho_A\equiv{\rm Tr}_{\bar A}|\psi\rangle\langle\psi|$ 
of  the subsystem $A$, \emph{i.e.} 
\begin{equation}
S_A\equiv-{\rm Tr}\boldsymbol\rho_A\ln\boldsymbol \rho_A.
\label{Sdef}
\end{equation}
The time evolution of the entanglement entropy after a quantum quench has been the focus of intense research  \cite{calabrese_evolution_2005,FC:XY,alba_entanglement_2016,dsvc17,ep-08,nr-14,kormos-2014,leda-2014,collura-2014,bhy-17,hbmr-17,dmcf-06,lauchli-2008,hk-13,fc-15,buyskikh-2016,kctc-17,coser-2014,cotler-2016,nahum-17,p-18,fnr-17,r-2017,daley-2012,mkz-17,evdcz-18,ckt-18, CC:lq, SD:lq, PE:ql, PE:ql2, mariolq, brunolq, ryulq,alba-inh,mbpc-17, BFPC18}. 
One of the main reasons for this fervent activity is technological: the rate of growth of the entanglement entropy determines whether it is feasible to simulate non-equilibrium quantum systems with tensor network algorithms~\cite{swvc-08,swvc-08b,pv-08,rev1,d-17}. Another reason was in relation to the key question on the 
emergence of thermodynamics from the microscopic quantum dynamics. There is now solid evidence indicating that the thermodynamic entropy in the local stationary state following the quench is nothing but the asymptotic entanglement entropy of a large subsystem~\cite{calabrese_evolution_2005,alba_entanglement_2016,dls-13,bam-15,Gur14,SPR11}; this is also supported by direct experimental evidence \cite{kaufman-2016}.

Despite years of intensive investigation, analytic \emph{ab-initio} results for the entanglement entropy are 
scarce even for free theories because of the intrinsic difficulties in evaluating Eq.  \eqref{Sdef}. 
Surprisingly, many qualitative features of the entanglement entropy evolution may be understood using a very simple physical picture proposed in Ref.~\cite{calabrese_evolution_2005}.
In this {\it quasiparticle picture}, the initial state produces {\it pairs} of excitations with opposite momentum. 
For $t>0$, the excitations with momentum $k$ move ballistically with group velocity $v(k)$. 
The spreading of entanglement \cite{calabrese_evolution_2005} and correlations \cite{cc-06,cc-07} 
is interpreted in terms of entangled particles emitted from the same point in space.
As the quasiparticles move far apart, larger regions of the system get entangled: 
at a given time, the entanglement entropy of the subsystem $A$ is due to those quasiparticles that,  
emitted from the same point in space, are shared between subsystem $A$ and its complement. 
Thus, for an interval $A$ of length $\ell$ embedded in an infinite one-dimensional system we have~\cite{calabrese_evolution_2005} 
\begin{equation}
\label{semi-cl}
S_A(t)= 2t\!\!\!\!\int\limits_{\!2|v(k)|t<\ell}\!\!\!\!{\rm d} k\, |v(k)|s(k)+
\ell\!\!\!\!\int\limits_{2|v(k)|t>\ell}\!\!\!\!{\rm d}k\, s(k)\,,
\end{equation}
where we weighted the 
quasiparticles with a factor  $s(k)$ which encodes the production rate of quasiparticles with momentum $\pm k$ and
their individual contribution to the entanglement entropy. 

Obviously this picture is strictly valid only for those models where all quasiparticles have an infinite lifetime, 
\emph{i.e.} in integrable models. It gives, however, important qualitative and quantitative information in many other circumstances as well, see, \emph{e.g.}, Refs.~\cite{kctc-17,ckt-18}.   
For integrable models, Eq.~\eqref{semi-cl} is expected to be asymptotically exact in the space-time scaling 
limit $t,\ell\to \infty$ with the ratio $t/\ell$ fixed. 
When a maximum quasiparticle velocity $v_M\ge |v(k)|$ exists (\emph{e.g.}, as 
a consequence of the Lieb-Robinson bound~\cite{lieb-1972}), Eq.~\eqref{semi-cl} predicts 
that for $t\le \ell/(2v_M)$, the entanglement  entropy grows linearly in time because the second term in~\eqref{semi-cl} vanishes. 
Conversely, for $t\gg\ell/(2v_M)$, only the second term is nonzero and $S_A(t)$ becomes extensive in the subsystem size, 
namely $S_A(t)\propto\ell$.

In order to turn~\eqref{semi-cl} into a quantitative prediction, one should be able to determine the two functions 
$s(k)$ and $v(k)$. In particular, $v(k)$ is naturally identified with the velocity of excitations over the stationary state 
\cite{alba_entanglement_2016}, which 
can be readily computed by Bethe ansatz~\cite{BEL:prl}. Computing $s(k)$ from first principles is, however, a formidable task 
even for free models (see \emph{e.g.} \cite{FC:XY}). A key observation of Ref.~\cite{alba_entanglement_2016} was that such an \emph{ab-initio} calculation is not necessary: one can conjecture the form of the function $s(k)$ by imposing 
 the equality between the entanglement and the thermodynamic entropy in the stationary state. One notes that for integrable models 
the stationary state is given by a generalised Gibbs ensemble (GGE), which takes into
account all the constraints imposed by the local and quasi-local integrals of motion~\cite{vidmar-2016, ef-16, quasilocal-16, rigolGGE}.
The determination of the thermodynamic entropy in the GGE is a standard equilibrium calculation and 
the entropy density in momentum space $s(k)$ can be easily read off. The resulting $s(k)$ is finally used as entry in 
the quasiparticle formula~\eqref{semi-cl} making it quantitative. Importantly, this prediction is obtained without
solving the complex many-body dynamics.

A crucial assumption behind the conjecture \eqref{semi-cl} is that the initial state acts as a source of {\it pairs} of quasiparticle excitations 
with opposite momentum. In Bethe-ansatz language this assumption may be viewed as a consequence of the requirement that only 
{\it parity-invariant  eigenstates} (as defined in \cite{dwbc-14,cd-12}) have nonzero overlap with the initial state. It has been recently shown 
(first for quantum field theories \cite{delfino-14, schuricht_quantum_2015} and then for lattice integrable models \cite{PPV:integrablestateslattice}) that only quenches 
originating from these initial states are compatible with some integrability requirements that strongly simplify the solution of the quench problem.
Indeed, so far, all the overlaps which are exactly known in interacting integrable models on the lattice, satisfy the parity invariant 
constraint~\cite{dwbc-14,cd-12,pc-14,pozsgay-14,bdwc-14,cd-14,fz-16,dkm-16}. Moreover, states with this structure give a good approximation of the initial state after mass and interaction quenches in the sinh-Gordon field theory~\cite{sotiriadis_zamolodchikov_2012, sotiriadis_boundary_2014, horvath_initial_2016, horvath_overlaps_2017}. %
States with nonzero overlap with generic eigenstates or with different families of eigenstates do, however, exist, 
and it is of fundamental importance to understand if and how \eqref{semi-cl} could generalise. 
A very useful playground where one can gain valuable insight into this question is represented by free models, 
since there quench problems can be readily solved even for initial states with a more general overlap structure.

In a recent paper \cite{previousepisode}, in the context of quantum quenches for Hubbard model with infinite repulsion, 
we constructed several families of initial states that have nonzero overlap with states formed by generic multiplets of quasiparticles and 
not only pairs. 
Analogous states can be constructed for free spinless fermions hopping on a one-dimensional line, and the time evolution from such states can be
 determined exactly.
This is precisely the goal of this paper: we study the dynamics of free fermions after quenches from states with no pair 
structure, focusing on the dynamics of entanglement entropy and its relation with the diagonal entropy.
We work out a suitable generalisation of \eqref{semi-cl} which shows a highly nontrivial structure: for example, we show that its form cannot be fixed solely by requiring that the stationary value of the entanglement entropy and the thermodynamic entropy be equal. Moreover, 
we show that the stationary value of the entanglement entropy is generically not related to the diagonal entropy. Although we focus on the von 
Neumann entropy \eqref{Sdef}, our findings apply to all  R\'enyi entanglement entropies as for free systems the complications 
encountered in defining them~\cite{ac-17b,ac-17c} are not present.

The manuscript is organised as follows. In Sections~\ref{sec:model} and \ref{sec:initialstates} we respectively present the model and the family of initial states considered, while in Section~\ref{sec:stationarystate} we determine the stationary state describing expectation values of local observables at infinite times after the quench. In Section \ref{sec:entanglemententropy} we analyse the time evolution of the entanglement entropy, and in Section~\ref{sec:diagonalvsthermodynamic} we study the relation between the stationary value of the entanglement entropy and the diagonal entropy. Finally, in Section~\ref{sec:conclusions} we report our conclusions. Three appendices complement the main text with a number of technical points.

\section{Model}
\label{sec:model}
We consider spinless fermions on the lattice, whose dynamics are described by the following Hamiltonian 
\begin{equation}
\hat H = -J\sum_{x=1}^{L}\left(c_{x}^\dagger c^{\phantom{\dagger}}_{x-1} + c_{x-1}^\dagger c^{\phantom{\dagger}}_{x}\right )\qquad\qquad c_{0}=c_{L}\,.
\label{eq:Hff}
\end{equation}
Here the fermionic operators $c_{x}$ and $c_{x}^\dagger$ satisfy the canonical anti-commutation relations
\begin{equation}
\{ c_{x}, c_{y} \} = \{ c_{x}^\dagger, c_{y}^\dagger \} = 0, \qquad\qquad
\{ c_{x}, c_{y}^\dagger \} = \delta_{x,y}\,.
\label{eq:CAR}
\end{equation}
We also introduce the unitary translation $\hat T$ and reflection $\hat R$ operators acting as follows on the fermionic operators 
\be
\hat Tc^\dagger_{x} \hat T^\dagger=c^\dagger_{x+1}\,,\qquad\qquad\qquad\hat Rc^\dagger_{x} \hat R^\dagger=c^\dagger_{L+1-x}\,.
\label{eq:translation-op}
\ee
The Hamiltonian \eqref{eq:Hff} is readily diagonalised by Fourier transform 
\be
\tilde c_k=\frac{1}{\sqrt{L}}\sum_{x=1}^L e^{i k x} c_x\,,\qquad\qquad k=\frac{2\pi}{L}n\,,\qquad\qquad n\in\mathbb{Z}\cap[-\tfrac{L}{2},\tfrac{L}{2})\,,
\ee
this transformation preserves the commutation relations 
\begin{equation}
\{\tilde c_{k}, \tilde c_{p} \} = \{\tilde c_{k}^\dagger, \tilde c_{p}^\dagger \} = 0, \qquad\qquad
\{\tilde c_{k}, \tilde c_{p}^\dagger \} = \delta_{k,p}\,.
\end{equation}
After the transformation the Hamiltonian reads as
\be
\hat H=\sum_k \varepsilon(k) \tilde c^\dag_k \tilde c^{\phantom{\dag}}_k\,,\qquad\qquad\qquad  \varepsilon(k)\equiv-2J\cos(k)\,.
\ee
Since 
\be
e^{i \hat H t} \tilde c^\dag_k e^{-i \hat H t}= \tilde c^\dag_k e^{i \varepsilon(k) t}\,,\qquad\qquad \hat T\tilde c^\dagger_{k} \hat T^\dagger=\tilde c^\dagger_{k} e^{i k}\,,
\ee
the operators $\tilde c^{{\dag}}_k\,(\tilde c^{\phantom{\dag}}_k)$ are interpreted as those creating (destroying) a quasiparticle of momentum $k$ and energy $\varepsilon(k)$. A basis of eigenstates of \eqref{eq:Hff} is naturally given by the ``scattering states'' of such quasiparticles
\be
\mathcal S = \{\ket{\Psi_{N}(k_1,\ldots, k_N)}\equiv \tilde c^\dag_{k_1}\ldots \tilde c^\dag_{k_N}\ket{0}:\quad N\in\mathbb{N},\quad k_i=\frac{2\pi}{L}n_i,\quad n_i\in\mathbb{Z}\cap[-\tfrac{L}{2},\tfrac{L}{2}),\quad n_i< n_{i+1},\ \forall i\}\,,
\label{eq:basis}
\ee
where $\ket{0}$ is the ``vacuum state'' such that 
\be
\tilde c_{k}\ket{0}=0 \qquad \forall k\,.
\ee
The state $\ket{\Psi_{N}(k_1,\ldots, k_N)}$ has energy and momentum given by  
\be
E(k_1\ldots k_N)=\sum_{i=1}^N\varepsilon(k_i)\,,\qquad\qquad P(k_1\ldots k_N)=\left[\sum_{i=1}^N k_i\right]\text{mod}\,2\pi\,.
\ee
In the thermodynamic limit, $\lim_{\rm th}$, when one sends the volume of the system to infinity keeping the particle density fixed, the momenta $k_i$ become continuous variables and the eigenstates are conveniently parametrised by their density
\be
\rho(k_i)={\textstyle\lim_{\rm th}}\frac{1}{L(k_i-k_{i+1})}\,,
\label{eq:rootdensity}
\ee  
called the root density. 

\section{Initial states}
\label{sec:initialstates}
Our goal in this paper is to study the time evolution generated by the Hamiltonian~\eqref{eq:Hff} on the following class of states 
\begin{align}
\ket{\Phi^{\nu}_{\{a_0,\ldots,a_{\nu-1}\}}} &=\prod_{j=1}^{L/\nu} \left(\sum_{m=0}^{\nu-1}  a_{m}  c^\dagger_{\nu j-m}
\right)\ket{0}\,,\qquad\qquad \sum_{j=0}^{\nu-1} |a_j|^2=1\,,
\label{eq:FFinitial2}
\end{align}
well-defined in finite volume $L$ such that $L/\nu$ is integer. These states  are invariant under translation of $\nu$ sites (up to a global phase) and are Gaussian, \emph{i.e.} Wick's theorem holds on these states. The latter property follows by observing that they can be written as vacuum states for an appropriately defined set of fermionic operators, related to $c_x,c_x^\dag$ by a linear transformation~\cite{molinari_notes_2017}. Such a set of fermionic operators is explicitly constructed in Appendix~\ref{app:Wick}.

Before describing further properties of these states, we report some simple examples to clarify their structure. Taking $\nu=3$ and denoting fermions with a closed circle and empty sites with an open circle, we begin with two examples where only one $a_i$ is nonzero
\begin{align}
\ket{\Phi^{3}_{\{1,0,0\}}} &= \prod_{j=1}^{L/3} c^\dagger_{3j}\ket{0}
= \ket{\fullmoon\fullmoon\newmoon \, \fullmoon\fullmoon\newmoon \, \fullmoon\fullmoon\newmoon\ldots}\\
\ket{\Phi^{3}_{\{0,1,0\}}} &= \prod_{j=1}^{L/3} c^\dagger_{3j-1}\ket{0}
= \ket{\fullmoon\newmoon\fullmoon \, \fullmoon\newmoon\fullmoon \, \fullmoon\newmoon\fullmoon\ldots}
\end{align}
In these cases, the state is comprised of repeated blocks of $\nu=3$ sites. Instead, if we take both $a_0=a_1=1/\sqrt{2}$, we obtain a sum of terms where each block of three either has a fermion in the last or the middle position
\begin{align}
\ket{\Phi^{3}_{\{1/\sqrt{2},1/\sqrt{2},0\}}} &=\frac{1}{2^{L/6}} \prod_{j=1}^{L/3} (c^\dagger_{3j}+c^\dagger_{3j-1})\ket{0}\nn
&=\frac{1}{2^{L/6}}\left[ \ket{\fullmoon\fullmoon\newmoon \, \fullmoon\fullmoon\newmoon \, \fullmoon\fullmoon\newmoon\ldots}\right.\nn
&+ \ket{\fullmoon\newmoon\fullmoon \, \fullmoon\fullmoon\newmoon \, \fullmoon\fullmoon\newmoon\ldots}
+ \ket{\fullmoon\fullmoon\newmoon \, \fullmoon\newmoon\fullmoon \, \fullmoon\fullmoon\newmoon\ldots}
+ \ket{\fullmoon\fullmoon\newmoon \, \fullmoon\fullmoon\newmoon \, \fullmoon\newmoon\fullmoon\ldots}+\ldots\nn
&+ \left.\ket{\fullmoon\newmoon\fullmoon \, \fullmoon\newmoon\fullmoon \, \fullmoon\fullmoon\newmoon\ldots} + \ldots + 
 \ket{\fullmoon\newmoon\fullmoon \, \fullmoon\newmoon\fullmoon \, \fullmoon\newmoon\fullmoon\ldots}\right] \, .
\end{align}

Note that these states are generically not reflection symmetric. Specifically, we have 
\be
\hat R \ket{\Phi^{\nu}_{
\{a_0,\ldots,a_{\nu-1}\}}}=
(-1)^{\frac{L}{2\nu}\left[\frac{L}{\nu}-1\right]} 
\ket{\Phi^{\nu}_{
\left\{ a_{\nu-1},\ldots, a_0 \right\} }}.
\ee
The overlaps between the initial states \eqref{eq:FFinitial2} and the eigenstates \eqref{eq:basis} read as 
\begin{align}
\braket{\Phi^{\nu}_{\{a_m\}}|\Psi_{N}(\boldsymbol k)}
=&\frac{\delta_{N,{L}/\nu}}{L^{N/2} }
\prod_{j=1}^N\!\left[{\sum_{\ell=0}^{\nu-1}a_\ell e^{-i\ell k_j}}\right]{\det}_{N}\{e^{i \nu   k_ab} \}_{a,b=1\ldots,N} \,.
\label{eq:initial2-overlapphi}
\end{align}
These overlaps impose a macroscopic number of constraints on the eigenstates contributing to the dynamics, specifically we must have 
\be
N=\frac{L}{\nu}\,, \qquad {k}_i- k_j\neq0\!\!\!\mod\frac{2\pi}{\nu} \qquad \forall\, i,j=1,\dots,\frac{L}{\nu}\,.
\label{eq:c2}
\ee
In words the second constraint means that for any given $k\in \frac{2\pi}{L}\left(\mathbb{Z}\cap[\tfrac{\nu-2}{2\nu}L,\tfrac{1}{2}L)\right)$ only one momentum in the set
\be
\left\{k, k-\frac{2\pi}{\nu}\, \ldots k-\frac{2\pi (\nu-1)}{\nu}\right\}
\ee 
can be occupied. Such constraint for $\nu=2$ produces a pair structure in the eigenstates contributing to the time evolution: for each particle with momentum $k\in\frac{2\pi}{L}\left(\mathbb{Z}\cap[\tfrac{\nu-2}{2\nu}L,\tfrac{1}{2}L)\right)$ there is a hole with momentum $k-\pi$ and \emph{vice versa}. For $\nu>2$ the eigenstates have no pair structure: there are correlated $\nu$-plets formed by $\nu-1$ holes and one particle. Note that, for $\nu>2$, such states are non-integrable according to the definition of Ref.~\cite{PPV:integrablestateslattice}.

\section{The post-quench stationary state}
\label{sec:stationarystate}

By studying time evolution from the states $\ket{\Phi^{\nu}_{\{a_m\}}}$, it is easy to verify that one-site translational invariance is restored at infinite times; a proof of this statement is reported in Appendix~\ref{app:translational}. As a consequence, the post-quench steady state is fully characterised by the mode occupations, \emph{i.e.} the expectation values of the conserved operators $c^\dag_k c^{\phantom{\dag}}_k$ on the initial state \eqref{eq:FFinitial2}. An explicit calculation gives 
\begin{align}
\braket{\Phi^{\nu}_{\{a_m\}}|\tilde c^\dag_k \tilde c^{\phantom{\dag}}_k|\Phi^{\nu}_{\{a_m\}}}&=\frac{1}{L}\sum_{n,m=1}^L e^{i k (m-n)} \braket{\Phi^{\nu}_{\{a_m\}}|c^\dag_{n} c^{\phantom{\dag}}_m|\Phi^{\nu}_{\{a_m\}}}=\frac{1}{\nu}\sum_{n,m=1}^{\nu} e^{i k (m-n)} \braket{\Phi^{\nu}_{\{a_m\}}|c^\dag_{n} c^{\phantom{\dag}}_{m}|\Phi^{\nu}_{\{a_m\}}}\notag\\
&= \frac{1}{\nu}\left(1+\sum_{n=1}^{\nu-1} e^{i k n} A_n^*+\sum_{n=1}^{\nu-1} e^{-i k n} A_n\right)= \frac{1}{\nu}\left(1+2\sum_{n=1}^{\nu-1}  |A_n| \cos(k n-{\rm arg}\left[A_n\right]) \right)\,,
\label{eq:modeocc}
\end{align}
where we defined
\be
A_n\equiv\sum_{m=n}^{\nu-1}a_{m}a^*_{m-n}\,.
\ee
The mode occupations \eqref{eq:modeocc} specify the representative macrostate characterising the expectation values of local observables in the thermodynamic limit at infinite times. Such a representative macrostate is composed by all the eigenstates in $\mathcal S$ whose momenta are distributed according to the root density  
\be
\rho_s(k)=\frac{1}{2\pi}\braket{\Phi^{\nu}_{\{a_m\}}|\tilde c^\dag_k \tilde c^{\phantom{\dag}}_k|\Phi^{\nu}_{\{a_m\}}}=\frac{1}{2\nu\pi}\left(1+2\sum_{n=1}^{\nu-1}  |A_n| \cos(k n-{\rm arg}\left[A_n\right]) \right)\,.
\label{eq:rhos}
\ee
This expression agrees with the Quench Action result~\cite{previousepisode}. We note that, as a consequence of the constraints \eqref{eq:c2}, the root density of the stationary state satisfies  
\be
\int_{-\pi}^\pi {\rm d}k\, \rho_s(k)=\frac{1}{\nu}\,,
\ee
\be
\rho_s(k) + \rho_s\left(k-\frac{2\pi}{\nu}\right) + \ldots+ \rho_s\left(k-\frac{2\pi(\nu-1)}{\nu}\right)=\frac{1}{2\pi}\,.
\ee

\section{Dynamics of Entanglement Entropy}
\label{sec:entanglemententropy}

In this section we focus on the dynamics of the entanglement entropy of an interval $A=[1,\ell]$ with respect to the rest of the chain, after a quench 
from the states \eqref{eq:FFinitial2}. The reduced density matrix we should plug into \eqref{Sdef} is then given by $\boldsymbol \rho_A(t)=\textrm{Tr}_{\bar A}[\boldsymbol\rho(t)]$, where 
\be
\boldsymbol\rho(t)=e^{-i \hat H t}\ket{\Phi^{\nu}_{\{a_m\}}}\bra{\Phi^{\nu}_{\{a_m\}}}e^{i \hat H t}\,.
\ee

\subsection{Simplifications in the free fermionic case}
Since the entanglement entropy stores information about all the quantum correlations in the system~\cite{CCrev}, and in free fermionic systems evolving from Gaussian states the only independent correlation functions are the two point functions, it is natural to expect $S_A(t)$ to be fully determined by the set of all two point functions. This is shown in Refs.~\cite{peschel2003,lrv-03,pe-09} for general Gaussian states and is explained in the following. We introduce a new set of fermionic operators, the Majorana fermions, defined as 
\be
a^x_n=c^\dag_n+c_n\,,\qquad a^y_n=i c_n-ic^\dag_n\,,\qquad\qquad\{a^\alpha_i,a^\beta_j\}=2\, \delta_{ij}\delta_{\alpha\beta}\,.
\ee 
Then, we introduce a $2L\times 2 L $ matrix $\Gamma(t)$ storing all the two point correlation functions of the Majorana fermions (and hence also of the fermions $c_j,c^\dag_j$). The matrix $\Gamma(t)$ is written in terms of $2\times2$ blocks as follows
\be
\Gamma(t)=
\begin{pmatrix}
[\Gamma^{(2)}(t)]_{1,1} & \ldots & [\Gamma^{(2)}(t)]_{1, {L}}\\
\vdots & \ddots & \vdots\\
[\Gamma^{(2)}(t)]_{{L}, 1} & \ldots & [\Gamma^{(2)}(t)]_{L, L}
\end{pmatrix}\,,
\ee
where
\be
\!\!{\left[\Gamma^{(2)}(t)\right]}_{n,m}=\delta_{n m}I_2-\begin{bmatrix}
 \mathrm{Tr}[\boldsymbol\rho(t) a_{n}^x a_{m}^x ] & \mathrm{Tr}[\boldsymbol\rho(t) a_{n}^x a_{m}^y ] \\
 \mathrm{Tr}[\boldsymbol\rho(t) a_{n}^y a_{m}^x ] & \mathrm{Tr}[\boldsymbol\rho(t) a_{n}^y a_{m}^y ] 
\end{bmatrix}\,,\qquad\qquad n,m=1,\ldots,L\,,
\ee 
and $I_2$ is the $2\times2$ identity matrix. We are now in a position to write the precise connection between $S_A(t)$ and the correlation functions stored in $\Gamma(t)$ for free fermionic systems evolving from Gaussian 
states~\cite{calabrese_evolution_2005, FC:XY,peschel2003,lrv-03}  
\be
S_A(t)= -{\rm Tr}\left[\left(\frac{I_{2\ell}-\Gamma_A(t)}{2}\right) \log\left(\frac{I_{2\ell}-\Gamma_A(t)}{2}\right)\right]\,.
\label{eq:entropy}
\ee
Here $I_{2\ell}$ is the $2\ell\times2\ell$ identity matrix and the ``reduced" correlation matrix $\Gamma_A(t)$ is obtained from $\Gamma(t)$ by taking only the first $2\ell$ rows and columns. This form drastically simplifies the calculation of the entanglement entropy: instead of the full reduced density matrix $\boldsymbol \rho_A(t)$, one only needs to consider fermionic two-point functions. 

A further simplification is achieved in our case by exploiting the $\nu$-sites translational invariance of the states $\ket{\Phi^{\nu}_{\{a_m\}}}$. In this case $\Gamma(t)$ acquires many useful mathematical properties that further simplify its explicit calculation, see \emph{e.g.} Appendix A of Ref.~\cite{BF15} for details. The result is conveniently expressed by arranging $\Gamma(t)$ in $2\nu\times 2 \nu$ blocks $[\Gamma^{(2\nu)}(t)]_{n,m}$ as follows 
\be
\Gamma(t)=
\begin{pmatrix}
[\Gamma^{(2\nu)}(t)]_{1,1} & \ldots & [\Gamma^{(2\nu)}(t)]_{1, \frac{L}{\nu}}\\
\vdots & \ddots & \vdots\\
[\Gamma^{(2\nu)}(t)]_{\frac{L}{\nu}, 1} & \ldots & [\Gamma^{(2\nu)}(t)]_{\frac{L}{\nu},\frac{L}{\nu}}
\end{pmatrix}\,.
\label{eq:corrmatblock}
\ee
Each block is explicitly computed in the thermodynamic limit as 
\be
{\textstyle\lim_{\rm th}}[\Gamma^{(2\nu)}(t)]_{n,m}=\int_{-\pi}^\pi\frac{{\rm d} k}{2\pi} e^{i k (n-m)} e^{-i \mathcal H^{(2\nu)}(k) t}\Gamma^{(2\nu)}(k)e^{i \mathcal H^{(2\nu)}(k) t}\,.
\label{eq:blocktimet}
\ee
Here $\mathcal H^{(2\nu)}(k)$ is a $2\nu\times2\nu$ matrix known as the $2\nu\times2\nu$ ``symbol" of the Hamiltonian \eqref{eq:Hff} (\emph{cf}. Appendix A of  \cite{BF15}) and reads as
\be
\mathcal H^{(2\nu)}(k)=-J\begin{pmatrix}
0 & \sigma^y & 0 & \ldots & \hspace{0.125cm}0\\
\sigma^y & 0  & \sigma^y  & &\hspace{0.125cm}\vdots\\
0 & \ddots &  \ddots& \ddots & \hspace{0.125cm}0 \\
\vdots & & \sigma^y & 0  &\hspace{0.125cm}\sigma^y \\
0 & \ldots & 0 & \sigma^y & \hspace{0.125cm}0
\end{pmatrix}-J \begin{pmatrix}
0 & 0 & \ldots & 0 & e^{-i k}\sigma^y\\
0 & 0  & 0  & & 0\phantom{\vdots}\\
\vdots & \ddots &  \ddots& \ddots & \vdots \\
0\phantom{\vdots} & & 0  & 0  &0 \\
e^{i k}\sigma^y & 0 & \ldots & 0 & 0
\end{pmatrix}\,,
\ee
while $\Gamma^{(2\nu)}(k)$ is a $2\nu\times2\nu$ matrix known as the symbol of the correlation matrix at time $t=0$. For the states $\ket{\Phi^{\nu}_{\{a_m\}}}$ it reads as
\be
\Gamma^{(2\nu)}(k)=\begin{pmatrix}
[\Gamma(\{a_i\})]_{1,1} & \ldots & [\Gamma(\{a_i\})]_{1, \nu}\\
\vdots & \ddots & \vdots\\
[\Gamma(\{a_i\})]_{\nu, 1} & \ldots & [\Gamma(\{a_i\})]_{\nu, \nu}
\end{pmatrix}\,,
\ee
where 
\be
[\Gamma(\{a_i\})]_{n m}= (a^{\phantom{*}}_{\nu-n} a^*_{\nu-m}-a^{\phantom{*}}_{\nu-m} a_{\nu-n}^* )I_2 +(\delta_{nm} - a^{\phantom{*}}_{\nu-n} a^*_{\nu-m}-a^{\phantom{*}}_{\nu-m} a_{\nu-n}^* )\sigma^y\,.
\ee 
The matrices $I_2$ and $\sigma^y$ appearing in the above formula are respectively the $2\times2$ identity matrix and the second Pauli matrix. 

\subsubsection{Numerical Evaluation of $S_A(t)$}
\label{sec:num}
Equation~ \eqref{eq:blocktimet} can be used for a very efficient numerical calculation of the entanglement entropy in the thermodynamic limit. We discretise the time interval of interest $\{t_0=0,t_1\ldots,t_{N-1},t_N=t\}$, and at each discrete time $t_i$ we define the correlation matrix $\Gamma_{A}(t_i)$ as follows 
\be
\Gamma_A(t_i)=
\begin{pmatrix}
[\Gamma^{(2\nu)}(t_i)]_{1,1} & \ldots & [\Gamma^{(2\nu)}(t_i)]_{1, \frac{\ell}{\nu}}\\
\vdots & \ddots & \vdots\\
[\Gamma^{(2\nu)}(t_i)]_{\frac{\ell}{\nu}, 1} & \ldots & [\Gamma^{(2\nu)}(t_i)]_{\frac{\ell}{\nu},\frac{\ell}{\nu}}
\end{pmatrix}\,,
\ee
where $[\Gamma^{(2\nu)}(t_i)]_{n,m}$ are computed using \eqref{eq:blocktimet}. We then diagonalise it and evaluate \eqref{eq:entropy}. This procedure is designed to work directly in the thermodynamic limit and can be carried out for very large subsystem sizes $\ell\sim500$ and large times $t\sim100$.  

\subsection{Semiclassical picture}

As first shown in Ref.~\cite{calabrese_evolution_2005}, the time evolution of the entanglement entropy from states with sub-extensive entanglement entropy  in the scaling limit 
\be
\ell\rightarrow\infty\,,\quad\quad t\rightarrow\infty \quad \text{with} \qquad \ell/t\quad\text{fixed}\,,
\label{eq:scalinglimit}
\ee
can be nicely interpreted by means of the semiclassical picture described in the introduction. Since it is based on the pair structure of the distribution of excitations, we denote the prediction \eqref{semi-cl} by $S^{\rm pair}_A(t)$, namely we write
\begin{align}
S^{\rm pair}_A(t) &= \int_{-\pi}^{\pi}\!\!\! \mathrm{d}k \Bigl[
2|v(k)|t\, s(k) \vartheta(\ell-2|v(k)|t) + \ell\, s(k)\vartheta(2|v(k)|t-\ell)
\Bigr].\label{eq:Spairs}
\end{align}
Here $\vartheta(x)$ is the Heaviside function and the ``weight" $s(k)$ is given by 
\begin{align}
s(k) = s_\text{YY}[\rhop_{s}](k)
= \frac{1}{2\pi}\log\frac{1}{2\pi} -\rhop_{s}(k)\log\rhop_{s}(k)-\left[\frac{1}{2\pi}-\rhop_{s}(k)\right]\log\left[\frac{1}{2\pi}-\rhop_{s}(k)\right]\,,
\label{eq:yangyang}
\end{align}
where $s_\text{YY}[\rho](k)$ is the Yang-Yang entropy density and $\rhop_{s}(k)$ is the root density~\eqref{eq:rhos} of the stationary state. Finally, the velocity $v(k)$ is the group velocity of elementary excitations over the stationary state, which in our case reads as 
\begin{equation}
v(k) = \varepsilon'(k)= 2J\sin(k)\,.
\label{eq:velocity}
\end{equation}
This formula was analytically proven for quenches within the XY model in Ref.~\cite{FC:XY} and was shown to work also for interacting integrable models in Ref.~\cite{alba_entanglement_2016}, where several quenches from low entangled initial states to the XXZ spin-$1/2$ chain were considered. Moreover, it was generalised to include the case of initial states with extensive entanglement entropy, at least for free systems~\cite{CEF:ising1, leda-2014, kormos-2014}. In all these cases, however, the distribution of excitations produced by the quench is characterised by the pair structure. This is true also for those states considered in Ref.~\cite{alba_entanglement_2016} for which the overlaps are not explicitly known~\cite{PPV:integrablestateslattice}. 

Here we use the efficient numerical procedure described in the above section to investigate whether this formula is able to describe the time evolution of the entanglement entropy from the states \eqref{eq:FFinitial2}. It is not at all clear that it would, since such states do not generically produce pairs of correlated excitations.

\subsubsection{States with $\nu=2$}

\begin{figure}[t]
\begin{tabular}{cc}
\includegraphics[width=0.5\textwidth]{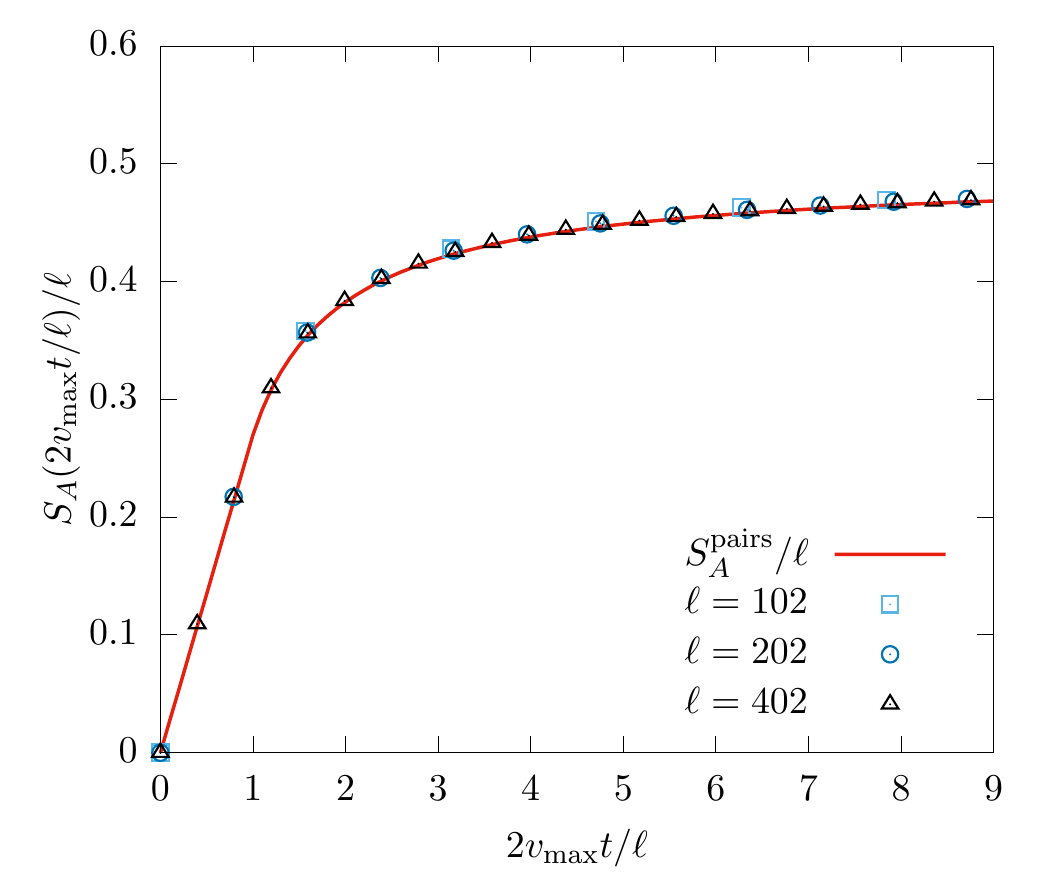} &\includegraphics[width=0.5\textwidth]{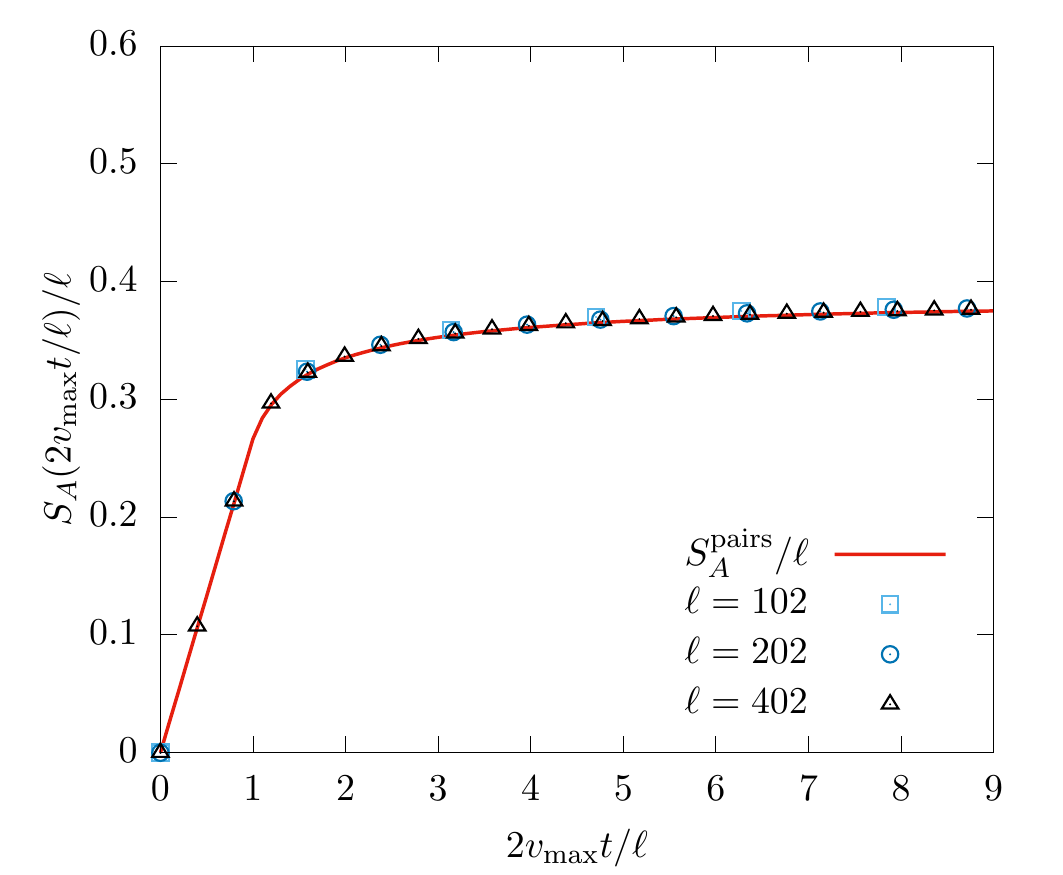}\\
\end{tabular}
\caption{Evolution of the entanglement entropy $S_A(t)/\ell$ in the rescaled time $2v_{\rm max} t/\ell$ ($v_{\rm max}=2 J$) for two different quenches from states $\ket{\Phi^{2}_{\{a_0,a_{1}\}}} $ and three different subsystem lengths $\ell$. The figure compares the results of the numerical calculations~\eqref{eq:entropy} (points) with the semiclassical prediction of Eq.~\eqref{eq:Spairs} (line). For the case reported on the left panel the parameters of the initial state are ${(a_0,a_1)=(0.88, 0.48 i)}$ while for that on the right panel  ${(a_0,a_1)=(1/\sqrt{2},e^{-i \frac{2 \pi}{10}}/\sqrt{2})}$.}
\label{Fig:pairstructure}
\end{figure}

Let us start considering states \eqref{eq:FFinitial2} with $\nu=2$. As noted in Section~\ref{sec:initialstates}, these states produce a distribution of excitations characterised by correlated particle-hole pairs with momenta $k$ and $k-\pi$. Even if now the excitations do not have opposite momenta, the semiclassical reasoning leading to \eqref{eq:Spairs} can be carried out with no modifications. This is because the two particles in the pair continue to have opposite velocities. We then expect  Eq.~\eqref{eq:Spairs} to correctly reproduce the evolution of entanglement entropy from the states $\ket{\Phi^{2}_{\{a_0,a_1\}}}$ in the scaling limit \eqref{eq:scalinglimit}. This is confirmed by our numerical simulations, as demonstrated in Fig.~\ref{Fig:pairstructure} for two representative examples. In the figure we plot $S_A(t)/\ell$ as a function of $2 v_{\rm max} t/\ell$, where we set $v_{\rm max}\equiv \max_k v(k)= 2 J$. The collapse of the data for increasing $\ell$ demonstrates that they have converged to the scaling limit.

\subsubsection{States with $\nu>2$}
\begin{figure}[t]
\includegraphics[width=0.48\textwidth]{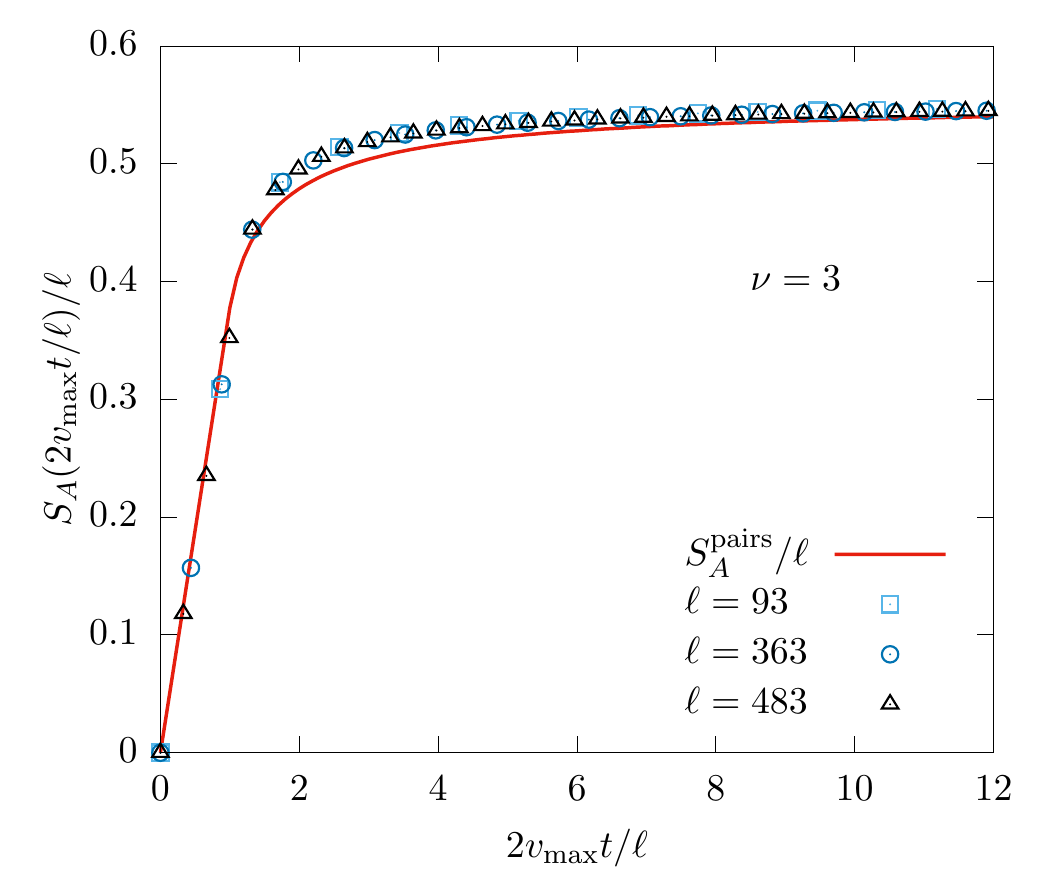}
\includegraphics[width=0.48\textwidth]{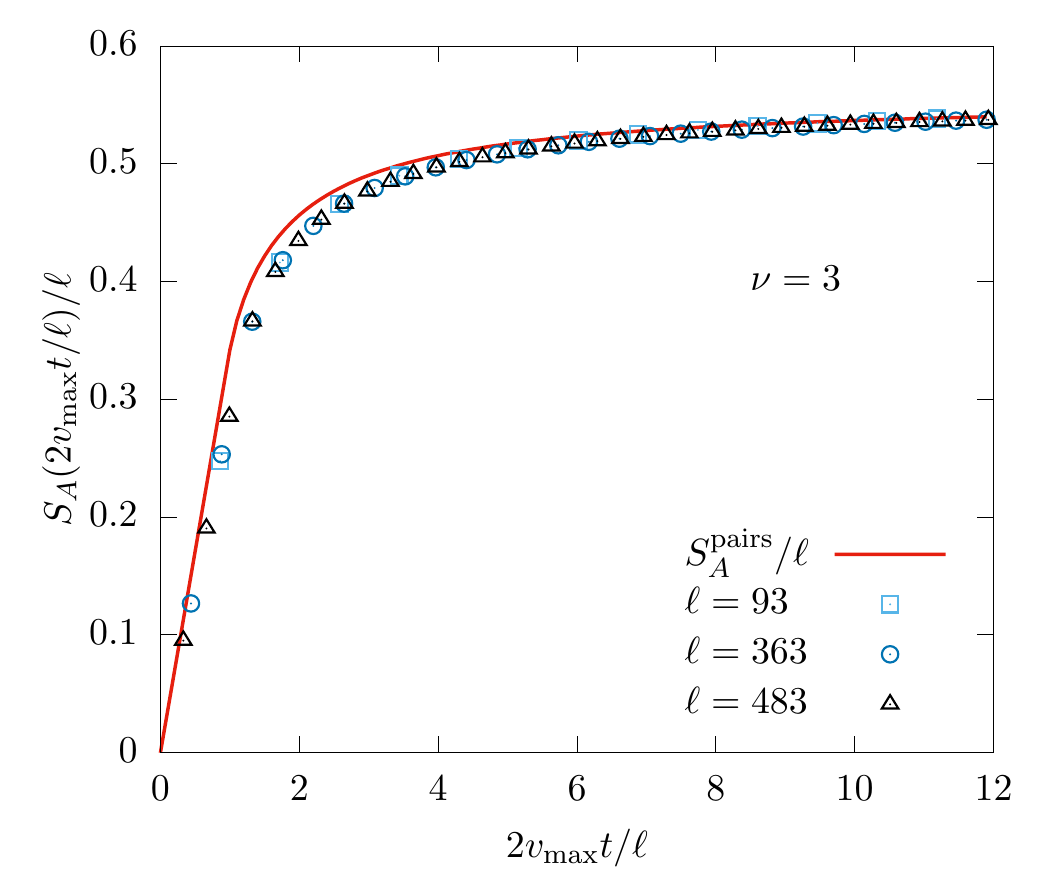}
\caption{Evolution of the entanglement entropy $S_A(t)/\ell$ in the rescaled time $2v_{\rm max} t/\ell$ ($v_{\rm max}=2 J$) after quenches from two different states $\ket{\Phi^{3}_{\{a_0,a_{1},a_2\}}} $ and three different subsystem-lengths $\ell$. The figure compares the results of the numerical calculations~\eqref{eq:entropy} with the semiclassical prediction of Eq.~\eqref{eq:Spairs}. For the case reported on the left panel the parameters of the initial state are ${(a_0,a_1,a_2)=(0.5,0.45\, e^{-i \frac{\pi}{6}}, 0.74\, e^{i \frac{5\pi}{7}})}$ while for that on the right panel  ${(a_0,a_1,a_2)=(0.12 , 0.46\,e^{-i \frac{\pi}{6}},0.88 e^{i\frac{35 \pi}{73}})}$.}
\label{Fig:nopairstructure}
\end{figure}

\begin{figure}[t]
\includegraphics[width=0.48\textwidth]{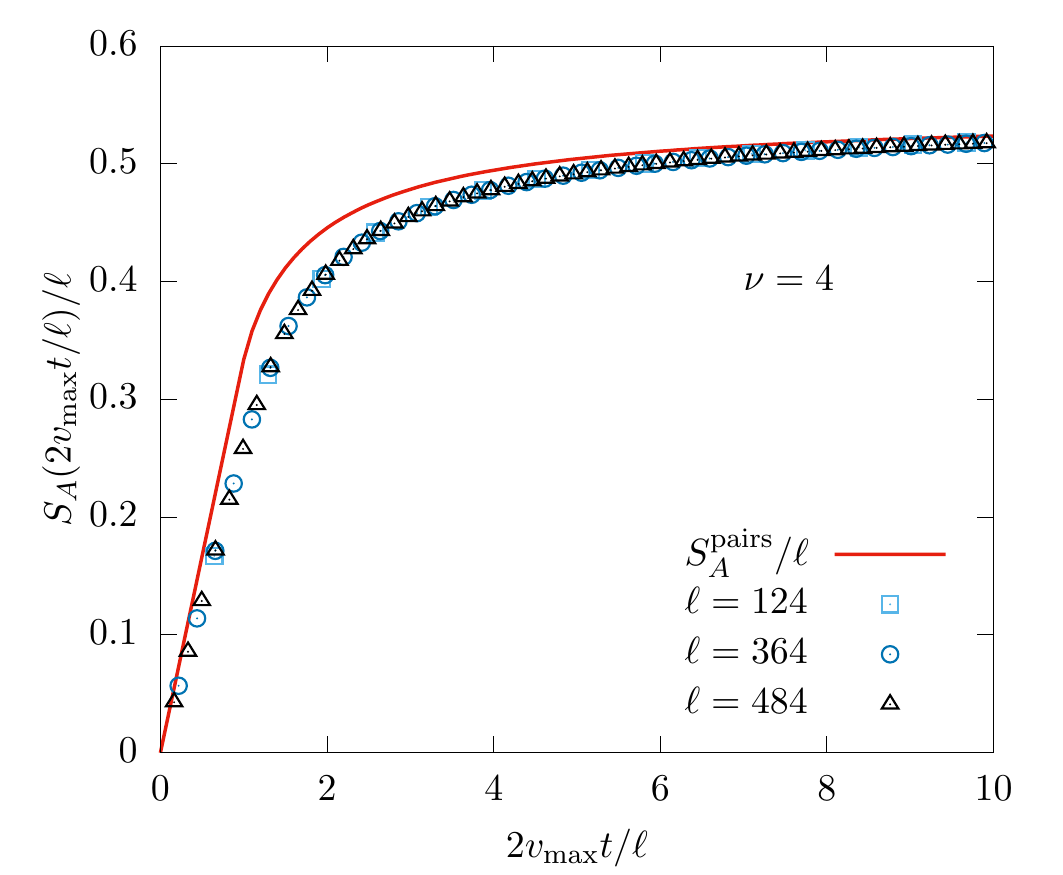}
\includegraphics[width=0.48\textwidth]{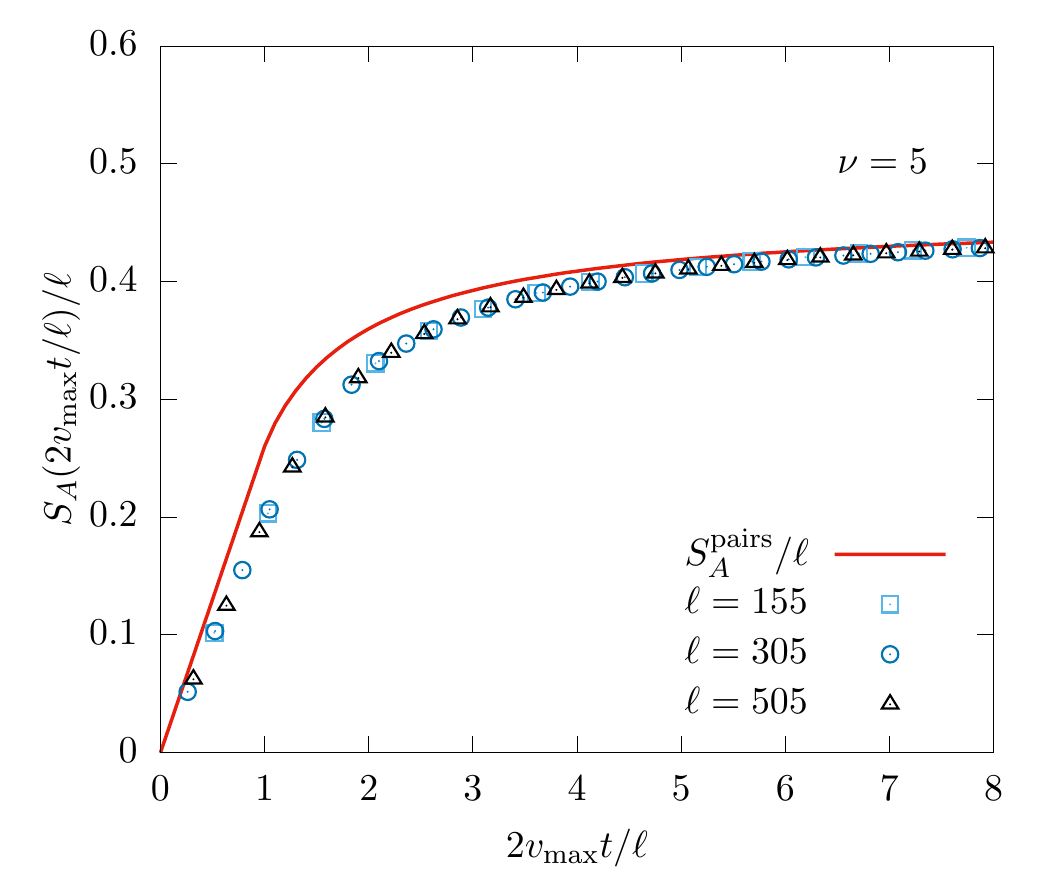}
\caption{Evolution of the entanglement entropy $S_A(t)/\ell$ in the rescaled time $2v_{\rm max} t/\ell$ ($v_{\rm max}=2 J$) after quenches from two different states $\ket{\Phi^{\nu}_{\{a_0,a_{1},a_2\}}} $ and three different subsystem-lengths $\ell$. The figure compares the results of the numerical calculations~\eqref{eq:entropy} with the semiclassical prediction of Eq.~\eqref{eq:Spairs}. For the case reported on the left panel the parameters of the initial state are $\nu=4$ and ${(a_0,a_1,a_2,a_3)=(0.11,0.44\, e^{i \frac{35\pi}{6}},0.85\, e^{i \frac{35\pi}{73}},0.26\, e^{i \frac{3\pi}{35}})}$ while for that on the right panel $\nu=5$ and ${(a_0,a_1,a_2,a_3,a_4)=(0.11 , 0.43\,e^{i \frac{5 \pi}{6}},0.25 \,e^{i\frac{3 \pi}{35}},0.82\,e^{i\frac{35 \pi}{73}},0.25\,e^{i\sqrt{\frac{3 \pi}{35}}})}$.}
\label{Fig:nopairstructurenuplets}
\end{figure}

Considering instead the case $\nu >2$, the situation becomes more complicated. As noted in Section~\ref{sec:initialstates}, such states impose a $\nu$-plet structure in the eigenstates of the Hamiltonian contributing to the dynamics instead of a pair one. Since there is no pair structure we have no reason to expect Eq.~\eqref{eq:Spairs} to hold. Our numerical calculations show that, in this case, Eq.~\eqref{eq:Spairs} does indeed not hold. This is demonstrated in Figs.~\ref{Fig:nopairstructure} and \ref{Fig:nopairstructurenuplets} for four representative examples. From these figures we clearly see that the numerical data for different $\ell$ have already reached convergence and that they show a distinct deviation from the prediction of Eq.~\eqref{eq:Spairs}. Note, however, that at infinite times the prediction of Eq.~\eqref{eq:Spairs} seems to be recovered. This is in agreement with the widely believed equality between entanglement entropy and thermodynamic entropy at infinite times after the quench~\cite{calabrese_evolution_2005,alba_entanglement_2016,dls-13,bam-15,Gur14,SPR11}. 

\subsubsection{Generalised semiclassical formula}

Our strategy here is to stick to the semiclassical picture and interpret the failure of Eq.~\eqref{eq:Spairs} as a consequence of the absence of the pair structure. Our goal is then to derive a semi-classical formula, which takes into account the more general $\nu$-plet structure in the excitations selected by the states \eqref{eq:FFinitial2} with $\nu>2$. To this aim it is convenient to introduce the following relabelling, which corresponds to a folding of the Brillouin zone. Instead of specifying the states using a single species of quasiparticles labelled by the momentum $k\in[-\pi,\pi]$ we use $\nu$ different species of quasiparticles with ``rapidity'' $p\in[\pi-2\pi/\nu,\pi]$. The dispersion relation of the new quasiparticles is given by  
\be
k_{j}(p)=p-\frac{2 (j-1) \pi}{\nu}\qquad\qquad\varepsilon_j(p)=2J\cos\left(p-\frac{2 (j-1) \pi}{\nu}\right)\qquad\qquad j=1,\ldots,\nu\,,
\ee
and their velocities
\be
v_{j}(p)=\frac{\partial \varepsilon_j(p)}{\partial k_j(p)}=2J\sin\left(p-\frac{2 (j-1) \pi}{\nu}\right)\qquad\qquad j=1,\ldots,\nu\,. 
\ee
With this relabelling the eigenstates of the Hamiltonian are described in the thermodynamic limit by the following set of root densities in rapidity space
\be
\rho^{(j)}(p)=\rho\left(p-\frac{2 (j-1) \pi}{\nu}\right)\qquad\qquad p\in\left[\pi-2\pi/\nu,\pi\right] \qquad\qquad j=1,\ldots,\nu\,,
\label{eq:constraintrho}
\ee
where $\rho(k)$ is the root density \eqref{eq:rootdensity}. In particular we denote by $\{\rho^{(j)}_s(p)\}_{j=1}^\nu$ the root densities of the stationary state~\eqref{eq:rhos}. The constraint \eqref{eq:c2} is reflected in the following condition on the root densities of the eigenstates with nonzero overlap 
\be
\sum_{j=1}^\nu \rho^{(j)}(p)=\frac{1}{2\pi}\,.
\label{eq:constraint}
\ee

To find a generalisation of \eqref{eq:Spairs} which is able to describe the $\nu>2$ case we make the following three assumptions
\begin{itemize}
\item[1.] The quench produces a $\nu$-plet of correlated excitations at every spacial point. 
\item[2.] Excitations move as free classical particles with velocities $v_j(p)$. 
\item[3.] Every time that a subset of the particles of a given $\nu$-plet is in the subsystem $A$ and the rest is out of it, there is a nontrivial contribution to the entanglement.  
\end{itemize} 
The only nontrivial step we need to take in order to get a testable prediction is to determine the contribution of a given partition of the $\nu$-plet to the entanglement. We will find it by generalising  \eqref{eq:yangyang}, valid in the case $\nu=2$. Let us first reconsider that case in order to develop a general strategy. We call $s_1(k)$ the contribution to the entanglement given by the pair of momentum $k$ with a particle of species $1$ within the system and its  companion of species $2$ out. We call $s_2(k)$ the contribution in the reversed case of the particle of species $2$ located in the system and that of species $1$ out of it. Eq.~\eqref{eq:Spairs} is obtained requiring
\be
s_1(k)=s_2(k)=s(k)=\frac{1}{2\pi}\log\frac{1}{2\pi} -\rhop^{(1)}_{s}(k)\log\rhop^{(1)}_{s}(k)-\left[\frac{1}{2\pi}-\rhop^{(1)}_{s}(k)\right]\log\left[\frac{1}{2\pi}-\rhop^{(1)}_{s}(k)\right]\,, \qquad k\in[0,\pi]\,.
\label{eq:contributionnu2}
\ee
Note that $s_1(k)=s_2(k)$ is in accordance with the basic property of bipartite entanglement: $S_A(t)=S_{\bar A}(t)$ where $\bar A$ is the complement of $A$.

A way to explain \eqref{eq:contributionnu2} is as follows. First we interpret $\rho^{(i)}_{s}(p)$ as the density of quasiparticle excitations of momentum $p$ and species $i$ produced by the quench. So that in a region of size $a$ 
\be
N^{(i)}(a;p)= a \rho^{(i)}_{s}(p){\rm d}p
\ee
excitations of species $i$ and momentum in $[p,p+{\rm d}p]$ are produced. Second, we consider a pair of momenta $p$ emitted from a point $x$ such that only the particle of species 1 is within the system at time $t$. In this case, all the quasiparticles of species $1$ and momentum in $[p,p+{\rm d}p]$ coming from a region $[x-\epsilon \ell /2,x+\epsilon \ell/2]$ are in the system at time $t$ for small enough $\epsilon$; those of species 2 are out of the system. This means that of the 
\be
N(p)= N^{(1)}(\epsilon \ell,p)+N^{(2)}(\epsilon \ell,p)\,,
\ee
particles of momentum $[p,p+{\rm d}p]$ produced in $[x-\epsilon \ell /2,x+\epsilon \ell/2]$, $N_{\rm in}(p)=N^{(1)}(\epsilon \ell, p)$ are in the system at time $t$ and $N_{\rm out}(p)=N^{(2)}(\epsilon \ell, p)$ are out of it. 

In this framework, the contribution to the entanglement entropy of the particles emitted from the region ${[x-\epsilon \ell /2,x+\epsilon \ell/2]}$ can be interpreted as the logarithm of the number of ways we can arrange $N(p)$ particles in two groups of $N_{\rm in}(p)$ and $N_{\rm out}(p)$, divided by the size of the region. In the scaling limit  \eqref{eq:scalinglimit}, which we denote by ${\textstyle \lim_{\rm sc}}$, we have 
\be
s_1(p){\rm d}p={\textstyle \lim_{\rm{sc}}} \frac{1}{\epsilon \ell}\log\frac{N(p)!}{N_{\rm in}(p)!N_{\rm out}(p)!}= s(p) {\rm d}p\,.
\label{eq:contribnu2sp1}
\ee 
The same reasoning can be carried out also assuming that particles of species $2$ are in the system at time $t$ and those of species $1$ are out: this is realised by exchanging $ \rho^{(1)}(p)$ and $\frac{1}{2\pi}-\rho^{(1)}(p)$ in the formulae above. Since \eqref{eq:contribnu2sp1} is symmetric under this exchange, we find
\be
s_1(p)=s_2(p)=s(p)\,,\qquad p\in[0,\pi]\,.
\ee 
We can proceed in the same way in the generic $\nu$ case. If only a subset of particles $\{j_1,...,j_m\}\subset \{1,\ldots,\nu\}$ of a given $\nu$-plet are in $A$ at time $t$, the numbers $N_{\rm in}(p)$ and $N_{\rm out}(p)$ read as 
\be
N_{\rm in}(p)= \epsilon \ell \left(\sum_{i=1}^m \rho_s^{(j_i)}(p)\right) {\rm d}p\,,\qquad\qquad N_{\rm out}(p)= \epsilon \ell  \left(\frac{1}{2\pi}-\sum_{i=1}^m \rho_s^{(j_i)}(p)\right) {\rm d}p\,. 
\ee
We then postulate that the contribution to the entanglement entropy is 
\begin{align}
s_{\{j_i\}}(p){\rm d}p&\equiv{\textstyle \lim_{sc}} \frac{1}{\epsilon \ell}\log\frac{N(p)!}{N_{\rm in}(p)!N_{\rm out}(p)!}\notag\\
& = \left[\frac{1}{2\pi}\log\frac{1}{2\pi} - \left[\sum_{i=1}^m \rho_s^{(j_i)}(p)\right]\log\left[\sum_{i=1}^m \rho_s^{(j_i)}(p)\right]- \left[\frac{1}{2\pi}-\sum_{i=1}^m \rho_s^{(j_i)}(p)\right]\log\left[\frac{1}{2\pi}-\sum_{i=1}^m \rho_s^{(j_i)}(p)\right]\right]{\rm d}p\,.
\label{eq:contribution}
\end{align} 
Note that, as a consequence of the constraint \eqref{eq:constraintrho}, this form is symmetric when exchanging $\{j_1,...,j_m\}$ with its complement ${\{1,\ldots,\nu\} \setminus \{j_1,...,j_m\}}$. This is again in accordance with the basic property of bipartite entanglement.

The assumptions 1., 2., and 3., supplemented with the form \eqref{eq:contribution} of the contribution, turn the problem of computing the time evolution of the entanglement entropy into a kinematic problem: we have to determine which particular bipartition of which particular $\nu$-plets contributes at each time. Before considering the problem for finite times, let us take the infinite time limit. In this limit only one particle for each correlated $\nu$-plet can be in $A$, as the particles have different velocity. This means that for infinite times the entanglement entropy reads as    
\be
S_A(\infty) = \ell \int_{\pi-\frac{2\pi}{\nu}}^\pi \sum_{j=1}^{\nu} s_{j}(p) =\ell \int_{-\pi}^{\pi}  s_\text{YY}[\rhop](k)\,.
\ee
Namely, the entanglement entropy at infinite times is given by the thermodynamic entropy, in agreement with the general expectations~\cite{Gur14,SPR11} and the results of Figure~\ref{Fig:nopairstructure}. This gives a nontrivial consistency check on the expression \eqref{eq:contribution}.   

\begin{figure}[t]
\includegraphics[width=0.48\textwidth]{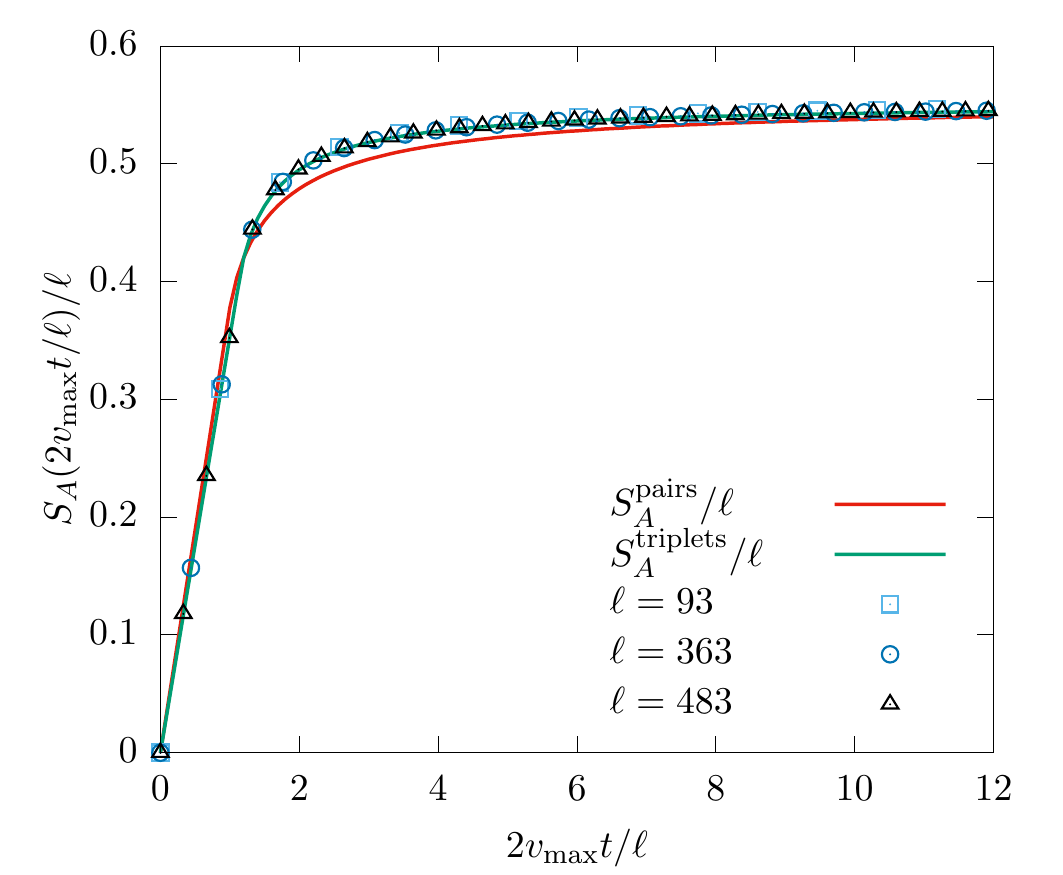}
\includegraphics[width=0.48\textwidth]{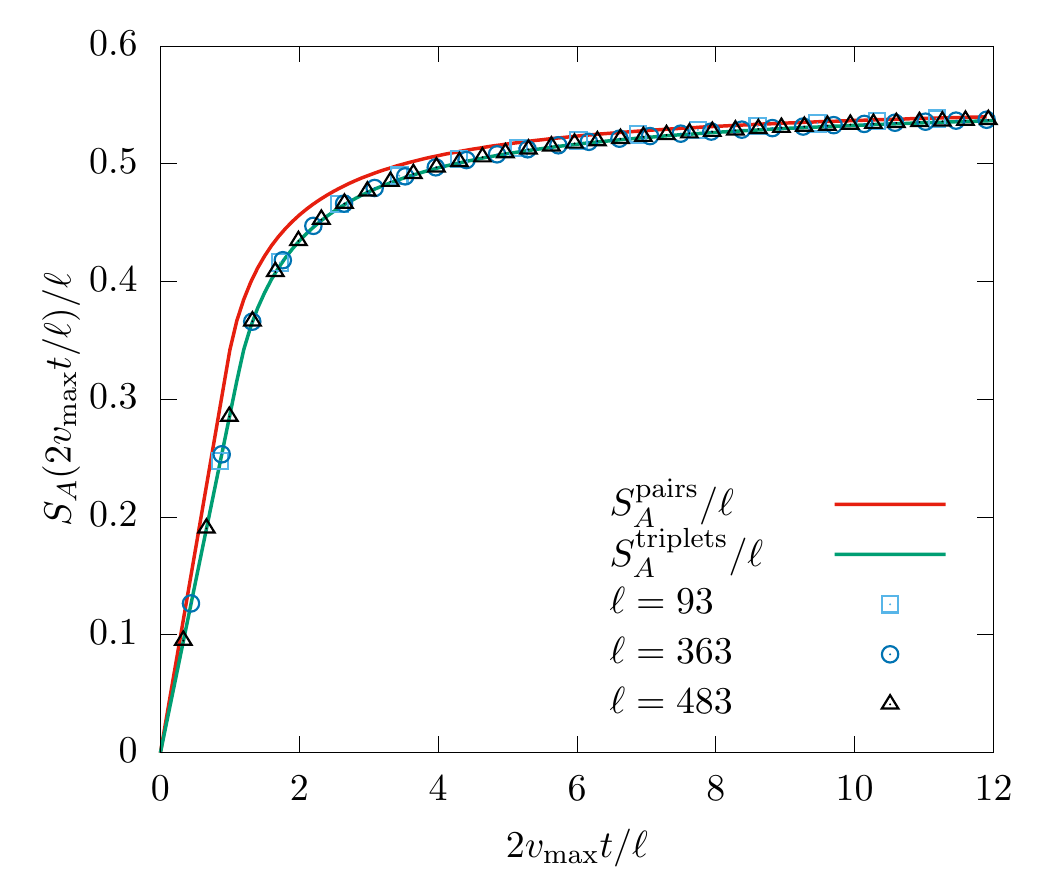}
\caption{Evolution of the entanglement entropy $S_A(t)/\ell$ in the rescaled time $2v_{\rm max} t/\ell$ (${v_{\rm max}=2J}$) after quenches from two different states $\ket{\Phi^{3}_{\{a_0,a_{1},a_2\}}} $ and three different subsystem-lengths $\ell$. The figure compares the results of the numerical calculations with the semiclassical predictions of Eq.~\eqref{eq:Spairs} and of Eq.~\eqref{eq:Striplets}. For the case reported on the left panel the parameters of the initial state are ${(a_0,a_1,a_2)=(0.5,0.45\, e^{-i \frac{\pi}{6}}, 0.74\, e^{i \frac{5\pi}{7}})}$ while for that on the right panel  ${(a_0,a_1,a_2)=(0.12 , 0.46\,e^{-i \frac{\pi}{6}},0.88 e^{i\frac{35 \pi}{73}})}$.}
\label{Fig:triplets}
\end{figure}

\begin{figure}[t]
\includegraphics[width=0.48\textwidth]{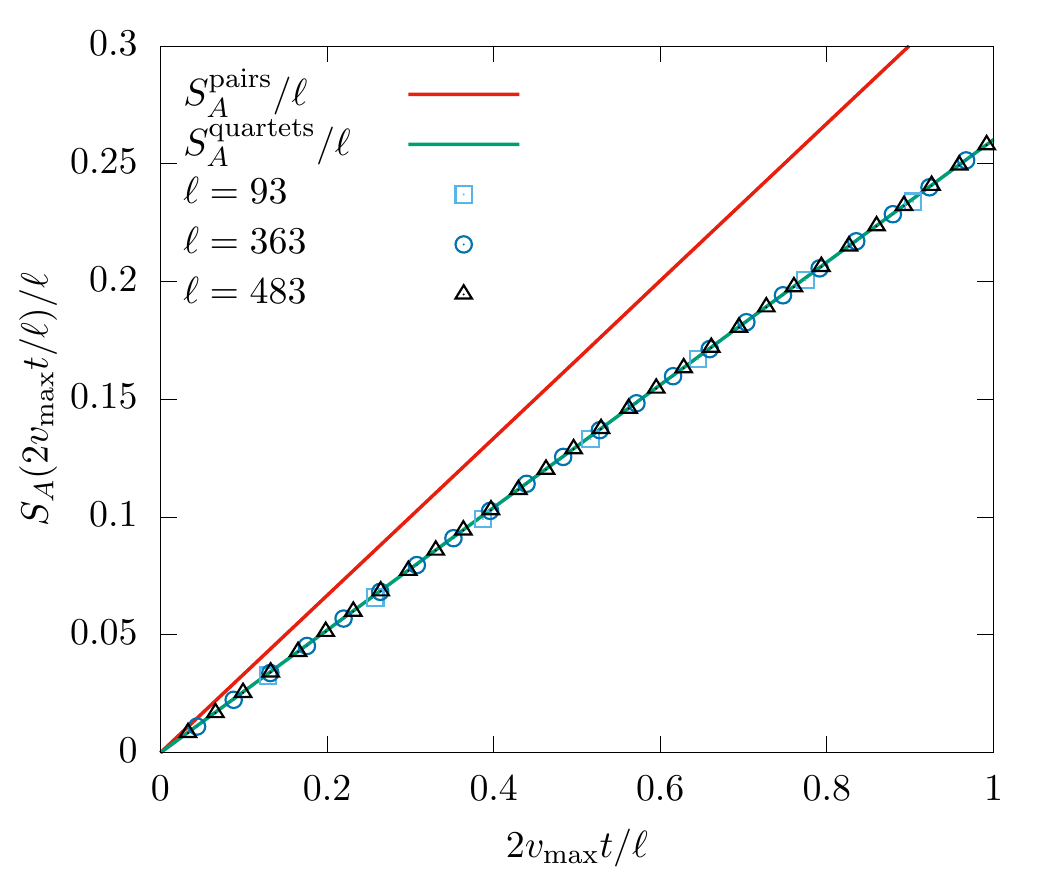}
\includegraphics[width=0.48\textwidth]{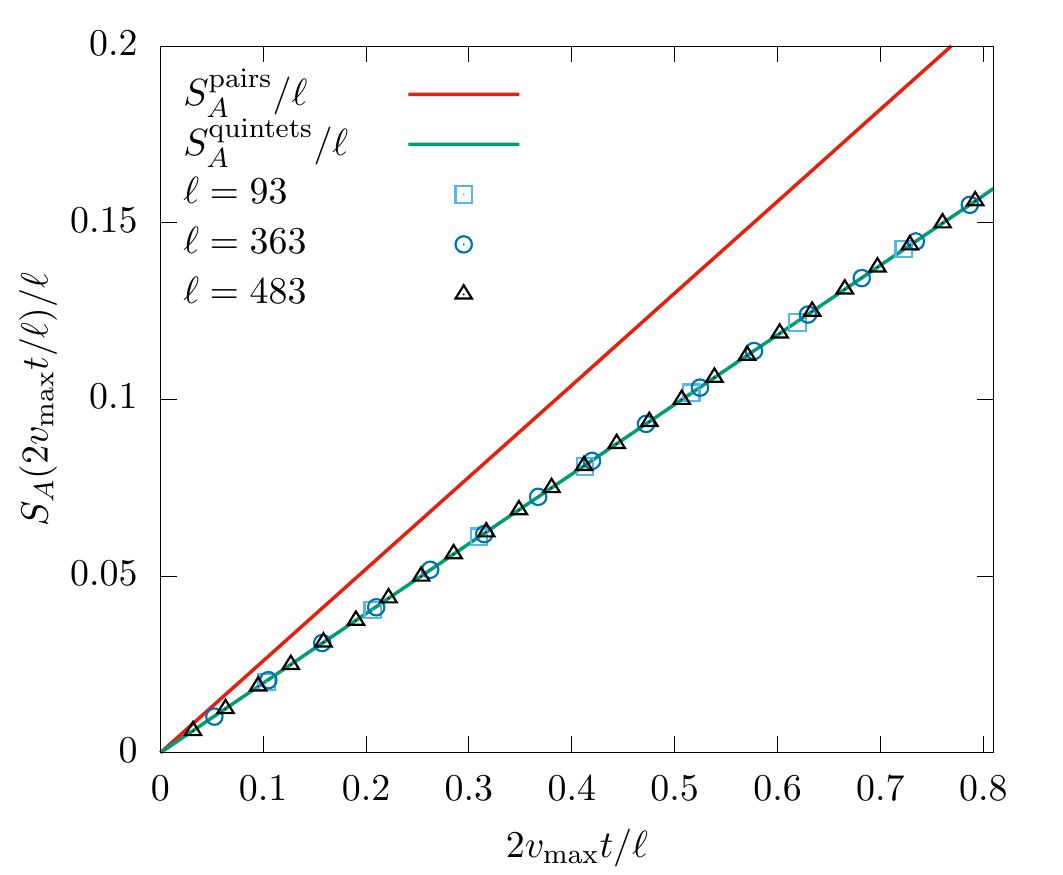}
\caption{Evolution of the entanglement entropy $S_A(t)/\ell$ in the rescaled time $2v_{\rm max} t/\ell$ (${v_{\rm max}=2J}$) after quenches from two different states $\ket{\Phi^{\nu}_{\{a_0,a_{1},a_2\}}} $ and three different subsystem-lengths $\ell$. The figure compares the results of the numerical calculations with the semiclassical predictions of Eq.~\eqref{eq:Spairs} and of Eq.~\eqref{eq:nuplets}, the latter is valid only for short times. For the case reported on the left panel the parameters of the initial state are $\nu=4$ and ${(a_0,a_1,a_2,a_3)=(0.11,0.44\, e^{i \frac{35\pi}{6}},0.85\, e^{i \frac{35\pi}{73}},0.26\, e^{i \frac{3\pi}{35}})}$ while for that on the right panel $\nu=5$ and ${(a_0,a_1,a_2,a_3,a_4)=(0.11 , 0.43\,e^{i \frac{5 \pi}{6}},0.25 \,e^{i\frac{3 \pi}{35}},0.82\,e^{i\frac{35 \pi}{73}},0.25\,e^{i\sqrt{\frac{3 \pi}{35}}})}$.}
\label{Fig:nuplets}
\end{figure}

Let  us now move to finding the entanglement entropy for finite times, for simplicity focusing on the case $\nu=3$. Let us consider a fixed $p$, since depending on the value of $p$ the velocities have a certain ordering. Let us consider the case $0<v_{3}(p)<v_{2}(p)<v_{1}(p)$~\cite{note}. We denote by
\be
S_A(t)|_{p,0<v_{3}(p)<v_{2}(p)<v_{1}(p)}
\ee
the contribution to the entanglement in this case.
\begin{figure}[t]
\includegraphics[width=0.8\textwidth]{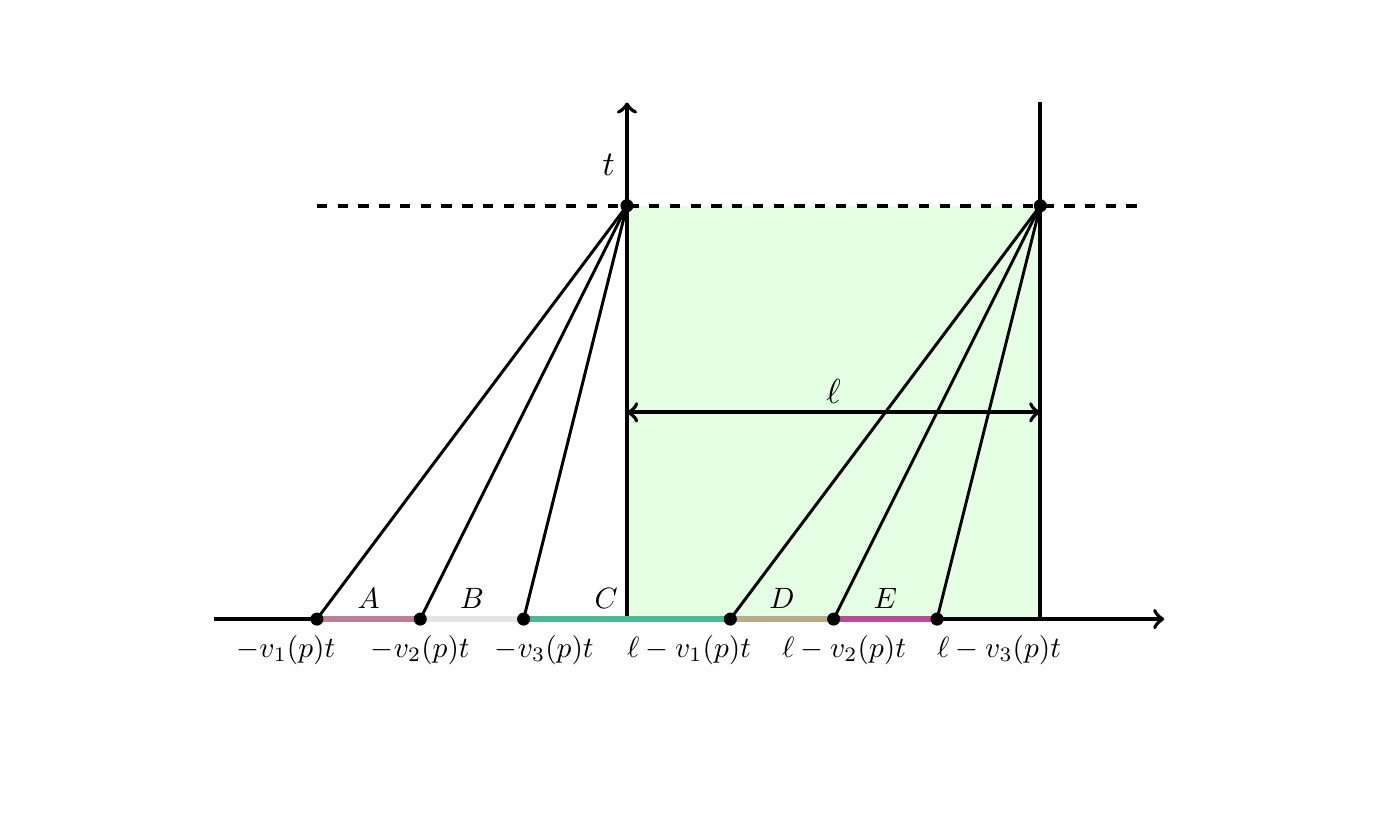}
\caption{Regions contributing to the entanglement entropy after a quench from a state $\ket{\Phi^{3}_{\{a_0,a_1,a_2\}}}$ for $0<v_{3}(p)<v_{2}(p)<v_{1}(p)$ and $t<\ell/(v_1(p)-v_3(p))$.}
\label{fig:qp}
\end{figure}
For short enough times $t$ the triplets contributing to the entanglement entropy are generated in the regions depicted in Fig.~\ref{fig:qp}. In particular 
\begin{itemize}
\item[-] No particles from the triplets generated on the left of $A$ or on the right of $E$ can be inside the system at time $t$ and hence contribute to the entanglement.
\item[-] Only particles of species $1$ from correlated triplets generated in the region $A$ can be inside the system at time $t$, so this region contributes with a term $s_{1}(p)(v_1(p)-v_2(p))t$. 
\item[-] Only particles of species 1 and 2 from the correlated triplets generated in the region $B$ are within the system, so this region gives a contribution of the form $s_{3}(p)(v_2(p)-v_3(p))t$.
\item[-] All the particles from the correlated triplets created in the region $C$ are within the system, so there is then no contribution to the entanglement from this region. 
\item[-] Only particles of species 2 and 3 from the triplets originating in $D$ are within the system, so we have the contribution $s_{1}(p)(v_1(p)-v_2(p))t$.
\item[-] Only particles of species 3 are within the system for the triplets originating in $E$, so we have $s_{3}(p)(v_2(p)-v_3(p))t$.
\end{itemize}
The situation changes when the time becomes equal to $\ell/(v_1(p)-v_3(p))$, since at that time the region $C$ shrinks to 0, and one has to consider a configuration different from that in Fig.~\ref{fig:qp}, where the regions $B$ and $D$ overlap. It is again very easy to find all the relevant contributions as we did above. Proceeding in this way, we find the entire contribution given by triplets of momentum $p$ can be written as  
\be
S_A(t)|_{p,0<v_3(p)<v_2(p)<v_1(p)} = f(\ell,t,p|3,2,1)\,,
\ee 
where we introduced the function 
\begin{align}
&f(\ell,t,p|a,b,c) \equiv \notag\\
&2 t \Bigl [s_{a}(p)(v_b(p)-v_a(p)) + s_c(p) (v_c(p) - v_b(p))\Bigr] \theta\left(\tau_{ca}(p)- t\right) \notag\\
&+ \Bigl[s_a(p) (\ell - (v_a(p)-2v_b(p)+v_c(p)) t ) + s_b(p) (((v_c(p)-v_a(p))t - \ell)   \notag\\
&\qquad + s_c(p) (\ell + (v_a(p)-2v_b(p) +v_c(p)) t)\Bigr]\chi_t\left(\left[\tau_{ca}(p),\min\left[\tau_{ba}(p),\tau_{cb}(p)\right]\right]\right)\notag\\
&+\theta(\tau_{cb}(p)-\tau_{ba}(p))\Bigl[s_a(p) (2\ell + (v_b(p) - v_c(p))t) + (s_b(p)+s_c(p)) (v_c(p) - v_b(p)) t \Bigr]\chi_t\left(\left[\tau_{ba}(p),\tau_{cb}(p)\right]\right)\notag\\
&+\theta(\tau_{ba}(p)-\tau_{cb}(p))\Bigl[s_c(p) (2\ell + (v_a(p)-v_b(p))t) + (s_b(p)+s_a(p)) (v_b(p)-v_a(p)) t \Bigr]\chi_t\left(\left[\tau_{cb}(p),\tau_{ba}(p)\right]\right)\notag\\
&+\ell \Bigl[s_a(p) + s_b(p) +s_c(p)\Bigr] \theta \left(t - \max\left[\tau_{ba}(p),\tau_{cb}(p)\right]\right)\,.
\end{align}
Here $\chi_x([a,b])$ is the characteristic function of the interval $[a,b]$ and we defined 
\be
\tau_{ij}(p)\equiv\frac{\ell}{v_i(p)-v_j(p)}\,.
\ee
In general, proceeding as above, we see that if the velocities have the ordering $v_{\sigma(3)}(p)<v_{\sigma(2)}(p)<v_{\sigma(1)}(p)$, where $\sigma$ is a generic permutation in $\mathcal S_3$, the contribution to the entanglement entropy is given by   
\be
S_A(t)|_{p,v_{\sigma(3)}(p)<v_{\sigma(2)}(p)<v_{\sigma(1)}(p)} = f(\ell,t,p|\sigma(3),\sigma(2),\sigma(1))\,.
\ee
We can then write the total entanglement entropy as follows 
\begin{align}
S^{\rm triplets}_A(t) =&  \sum_{\sigma\in \mathcal S_3}\, \int_{\pi/3}^\pi \!\!\!{\rm d}p\,\,\theta(v_{\sigma(3)}<v_{\sigma(2)}(p)<v_{\sigma(1)}(p))\,\,f(\ell,t,p|\sigma(3),\sigma(2),\sigma(1))\,.
\label{eq:Striplets}
\end{align}
In Fig.~\ref{Fig:triplets} we compare the prediction of \eqref{eq:Striplets} with the numerical calculations for increasing $\ell$ in two representative examples. As we see from the plot, the agreement is extremely convincing. 
The explicit expression for generic $\nu$ is obtained reasoning in the same way, but it rapidly becomes quite cumbersome as there are many possible orderings for the $\tau_{ij}(p)$. Here we give the result in the simplifying limit of $\ell\rightarrow\infty$, \emph{i.e.}, when the system $A$ corresponds to the positive real half line
\begin{align}
\lim_{\ell\rightarrow\infty}S^{\nu{\text{-plets}}}_A(t) =&  \sum_{\sigma\in \mathcal S_\nu}\, \int_{\pi-2\pi/\nu}^\pi \!\!\!{\rm d}p\,\,\theta(v_{\sigma(\nu)}<\dots<v_{\sigma(1)}(p))\,\,\lim_{\ell\rightarrow\infty} f(\ell,t,p|\sigma(\nu),\ldots,\sigma(1))\,,
\label{eq:nuplets}
\end{align}
where 
\be
\lim_{\ell\rightarrow\infty} f(\ell,t,p|a_\nu,\ldots,a_1)=2 t\sum_{j=1}^{\nu-1} s_{\{a_1 \dots a_j\}}(p) (v_{a_j}(p) - v_{a_{j+1}}(p)) \,.
\ee
Formula \eqref{eq:nuplets} describes the behaviour of the entanglement entropy even at finite $\ell$, for times $t<\min_{ij}\min_p{\tau_{ij}(p)}$. In Fig.~\ref{Fig:nuplets} we compare the prediction of \eqref{eq:nuplets} with the numerical simulations, the agreement is again excellent and gives a further confirmation of the validity of Eq.~\eqref{eq:contribution}. 

\section{Diagonal vs Thermodynamic Entropies}
\label{sec:diagonalvsthermodynamic}

In this section we investigate the relation between the stationary values reached by the entanglement entropy, coinciding with those of the thermodynamic entropy of the stationary state~\eqref{eq:rhos}, and the recently proposed \emph{diagonal entropy}~\cite{P:diagonalentropy}. The latter quantity is regarded as an alternative microscopic definition of the entropy of a quantum system out of equilibrium and is given by the von Neumann entropy of the diagonal ensemble, namely  
\be
S_{\rm d} = -{\rm Tr}\boldsymbol \rho_{\rm d} \log \boldsymbol \rho_{\rm d}\,,
\label{eq:diagentr}
\ee
where 
\be
\boldsymbol \rho_{\rm d}=\sum_{n=0}^\infty\sum_{k_1<\ldots<k_n} |\braket{\Psi_{n}(k_1,\ldots,k_n)|\Phi^{\nu}_{\{a_m\}}}|^2 \ket{\Psi_{n}(k_1,\ldots,k_n)}\bra{\Psi_{n}(k_1,\ldots,k_n)}\,,
\ee
and $\{\ket{\Psi_{n}(k_1,\ldots,k_n)}\}$ are the eigenstates \eqref{eq:basis} of the Hamiltonian. The quantity~\eqref{eq:diagentr} has the important advantage of being time independent: it can be calculated on the initial state without solving the intricate many-body dynamics. The relation between $S_{\rm d}$ and the thermodynamic entropy  $S_{\rm th}$ has recently attracted considerable attention~\cite{leda-2014,kormos-2014, SPR11, collura-2014, Gur14,F:finitesize, PVCR17,SPR:pre,P:diagonalentropy,D:escort,ac-17b}. While it has been suggested that these two quantities are equivalent for generic systems~\cite{P:diagonalentropy}, all the cases examined in the context of integrable models revealed that the diagonal entropy was half of the thermodynamic one~\cite{leda-2014,kormos-2014, collura-2014, Gur14,F:finitesize, PVCR17,SPR:pre,D:escort}. The latter fact is connected with the pair structure in the eigenstates contributing to the dynamics: if in an integrable model the relevant eigenstates have the pair structure then the diagonal entropy is half of the thermodynamic one, as recently shown in Ref.~\cite{ac-17b}. Here we show that in the absence of the pair structure the ratio of $S_{\rm d}$ and  $S_{\rm th}$ is generically different from one and depends on the details of the initial state. Specifically, while for $\nu=2$ the ratio is fixed to $1/2$ in agreement with the result of Ref.~\cite{ac-17b}, for $\nu>2$ the ratio depends on the details of the states, \emph{i.e.}, on the configuration of $\{a_i\}$. We identify the range in which the ratio varies determining a maximum and a minimum value, $r_{\rm max}(\nu)$ and $r_{\rm min}(\nu)$, as a function of $\nu$. Moreover, we show that $1/2<r_{\rm min}(\nu)<r_{\rm max}(\nu)< 1$ for any finite $\nu$ and $\lim_{\nu\rightarrow\infty}r_{\rm max}(\nu)=1$, while $r_{\rm min}(\nu)$ appears to approach a constant $c\approx 0.53$.

\begin{figure}
\includegraphics[scale=0.8]{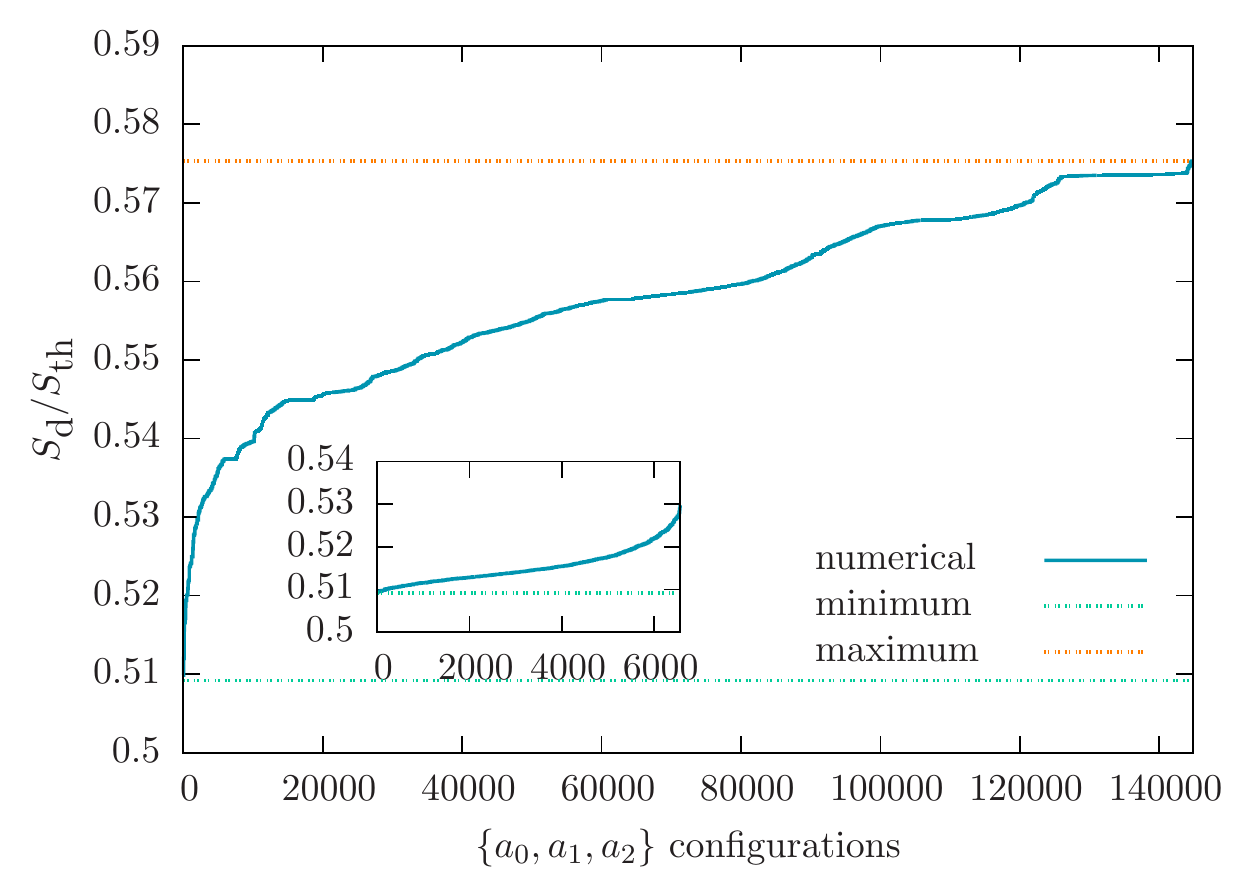}
\caption{Here we give numerical values for the ratio of the diagonal and thermodynamic entropies in the steady state \eqref{eq:rhos} for $\nu=3$. We parametrised the complex constants $a_i$ as in equation \eqref{eq:parametrisation}. The larger plot evaluates the ratio for values of the $a_i$ obtained by taking all combinations of $\theta_i=0,\pi/20,\ldots,\pi/2$,  $\alpha_i=0,\pi/20,\ldots,39\pi/20$ and $i=1,2$ which give unique ordered sets $\{a_0,a_1,a_2\}$. To demonstrate values closer to the minimum, we took a finer mesh around the values of the minimum: all combinations for $\theta_i=\theta_i^{\text{min}}-0.1$,  $\theta_i^{\text{min}}-0.075$, \ldots, $\theta_i^{\text{min}}+0.1$ and similarly for the $\alpha_i$. The dashed maximum value is given by the formula \eqref{eq:ratio-max} and the minimum value in Table \ref{tab:ratio-min}.}
\label{fig:entropy-ratio-numerics-nu-3}
\end{figure}

\begin{figure}
\includegraphics[scale=0.8]{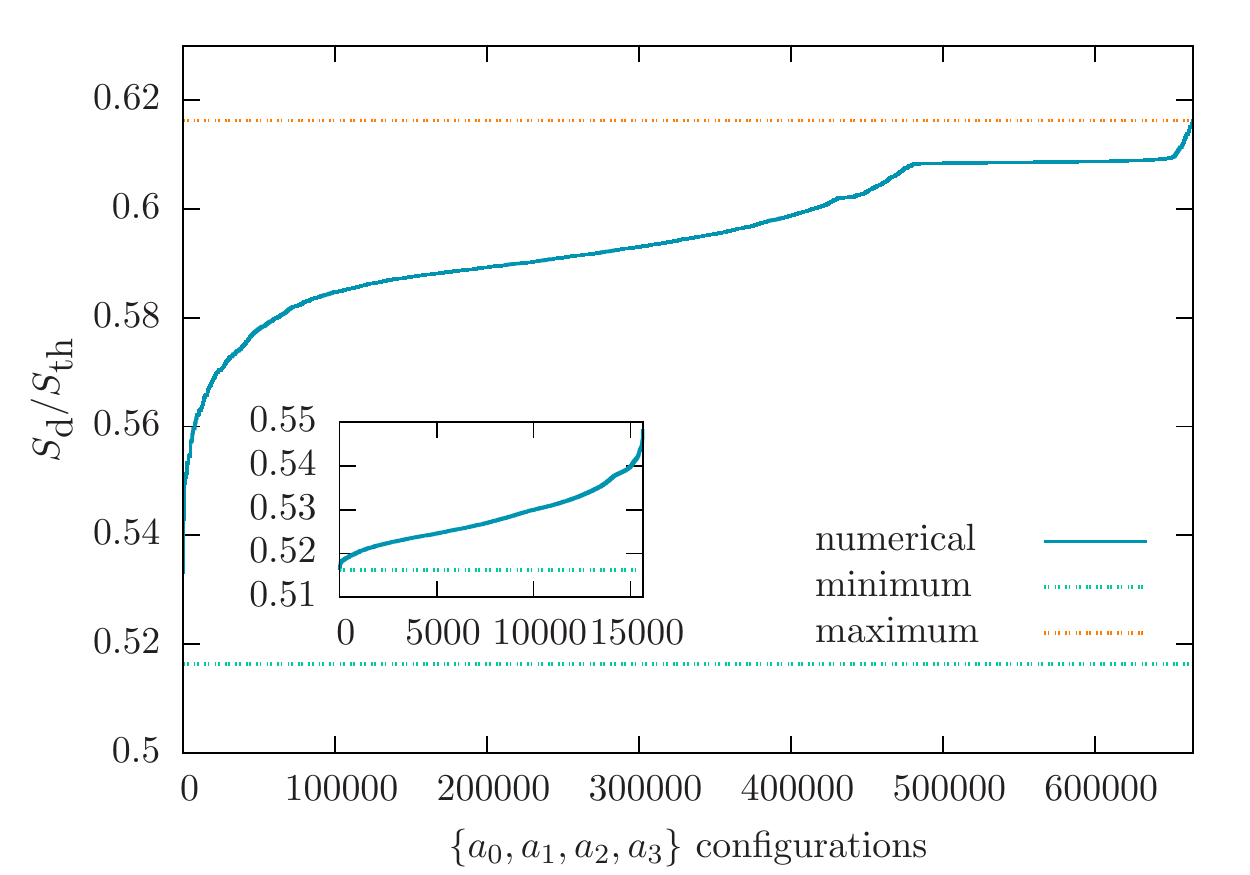}
\caption{Here we give numerical values for the ratio of the diagonal and thermodynamic entropies in the steady state \eqref{eq:rhos} for $\nu=4$. We parametrised the complex constants $a_i$ using polar coordinates for the modulus (in terms of angles $\theta_i$) and denoting the argument as $\alpha_i$. The larger plot evaluates the ratio for values of the $a_i$ obtained by taking all combinations of $\theta_i=0,\pi/10,\ldots,\pi/2$,  $\alpha_i=0,\pi/10,\ldots,19\pi/10$ and $i=1,2,3$ which give unique ordered sets $\{a_0,a_1,a_2,a_3\}$. To demonstrate values closer to the minimum, we took a finer mesh around the values of the minimum: all combinations for $\theta_i=\theta_i^{\text{min}}-0.1$, $\theta_i=\theta_i^{\text{min}}-0.05$, \ldots, $\theta_i^{\text{min}}+0.1$ and similarly for the $\alpha_i$. The dashed maximum value is given by the formula \eqref{eq:ratio-max} and the minimum value in Table \ref{tab:ratio-min}.}
\label{fig:entropy-ratio-numerics-nu-4}
\end{figure}

\begin{figure}
\includegraphics[scale=0.8]{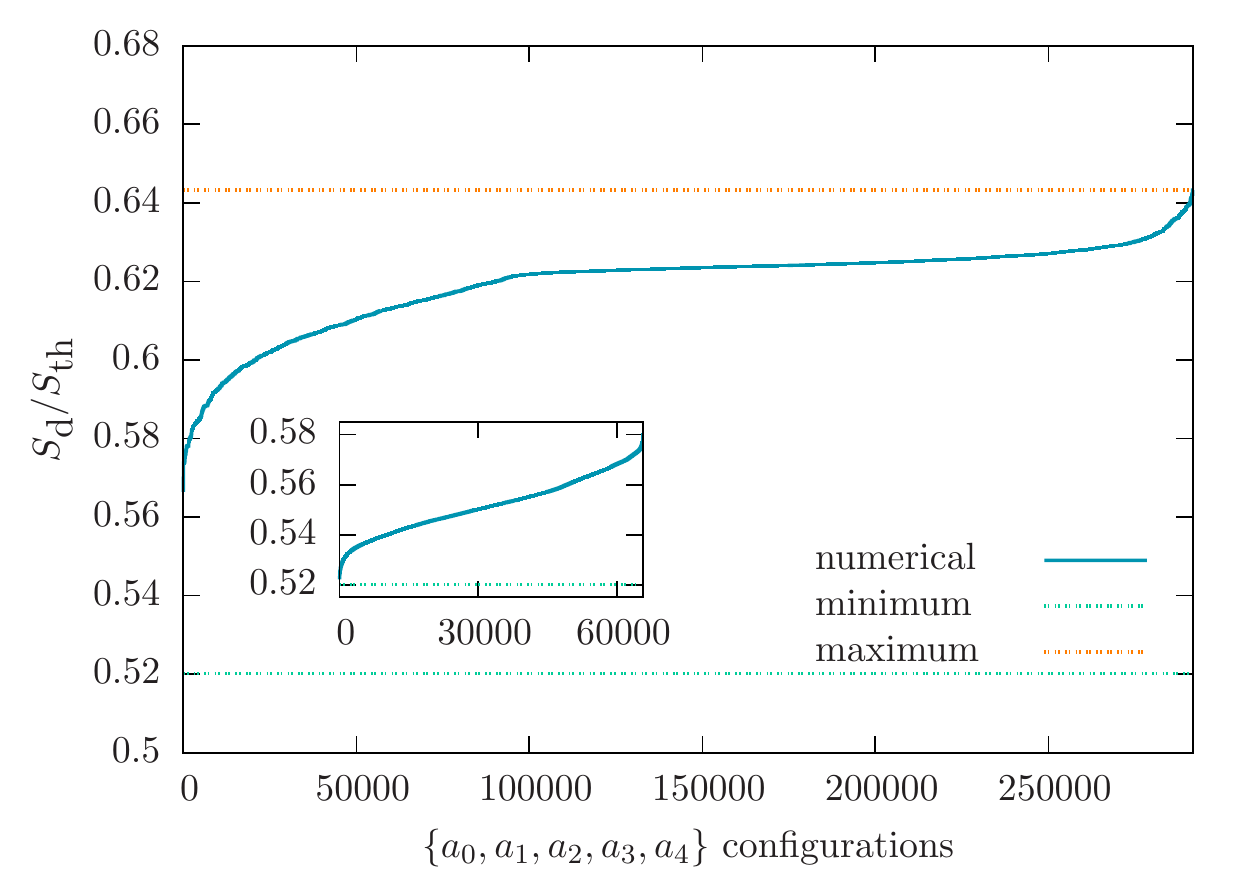}
\caption{Here we give numerical values for the ratio of the diagonal and thermodynamic entropies in the steady state \eqref{eq:rhos} for $\nu=5$. We parametrised the complex constants $a_i$ using polar coordinates for the modulus (in terms of angles $\theta_i$) and denoting the argument as $\alpha_i$. The larger plot evaluates the ratio for values of the $a_i$ obtained by taking all combinations of $\theta_i=0,\pi/6,\pi/3,\pi/2$,  $\alpha_i=0,\pi/5,\ldots,9\pi/5$ and $i=1,2,3,4$ which give unique ordered sets $\{a_i\}_{i=0}^4$. To demonstrate values closer to the minimum, we took a finer mesh around the values of the minimum: all combinations for $\theta_i=\theta_i^{\text{min}}-0.15$, $\theta_i^{\text{min}}-0.05$, $\theta_i^{\text{min}}+0.05$, $\theta_i^{\text{min}}+0.15$  and similarly for the $\alpha_i$. The dashed maximum value is given by the formula \eqref{eq:ratio-max} and the minimum value in Table \ref{tab:ratio-min}.}
\label{fig:entropy-ratio-numerics-nu-5}
\end{figure}

We begin by expressing the densities of thermodynamic entropy and diagonal entropy in the thermodynamic limit in terms of the root densities~\eqref{eq:constraintrho} of the stationary state. The thermodynamic entropy density is given by the integral of the Yang-Yang entropy. In terms of $\{\rho^{(j)}(p)\}_{j=1}^\nu$ (\emph{cf}.~\eqref{eq:constraintrho}) it reads as  
\be
\SYY[\{\rho^{(j)}\}]= \log \frac{1}{2\pi} - \sum_{j=1}^\nu \int_{\pi-\frac{2\pi}{\nu}}^\pi \mathrm{d}k \left\{ \rhop^{(j)}(k)\log\rhop^{(j)}(k)+\left[\frac{1}{2\pi}-\rhop^{(j)}(k)\right]\log\left[\frac{1}{2\pi}-\rhop^{(j)}(k)\right] \right\}\,.
\label{eq:yang-yang-entropy}
\ee
As shown in~\cite{ac-17b}, the thermodynamic limit of the diagonal entropy density is given by the reduced entropy appearing in the Quench Action approach~\cite{QA,caux_qareview} and defined as the Yang-Yang entropy density reduced to the states having nonzero overlap with the initial state. In our case, the reduced entropy was determined in~\cite{previousepisode} and reads as 
\be
\Sred[\{\rho^{(j)}\}] = 
\frac{1}{\nu}\log \frac{1}{2\pi} - \sum_{j=1}^\nu \int_{\pi-\frac{2\pi}{\nu}}^\pi \mathrm{d}k\, \rhop^{(j)}(k)\log\rhop^{(j)}(k)\,.
\label{eq:yangyangreduced}
\ee
Our quantity of interest can then be written as 
\begin{align}
r(\nu;\{a_i\}) &\equiv \frac{S_{\rm d}}{S_{\rm th}}=\frac{\Sred[\{\rho^{(j)}_s\}]}{\SYY[\{\rho^{(j)}_s\}]} \notag\\
&= \frac{
\frac{1}{\nu}\log\frac{1}{2\pi} - \int_{\pi-\frac{2\pi}{\nu}}^\pi \mathrm{d}k \sum_{j=1}^\nu\rho^{(j)}_s(k)\log \rho^{(j)}_s(k)
}{
\log\frac{1}{2\pi} - \int_{\pi-\frac{2\pi}{\nu}}^\pi \mathrm{d}k \sum_{j=1}^\nu
\left[\rho^{(j)}_s(k)\log \rho^{(j)}_s(k)
+(\frac{1}{2\pi}-\rho^{(j)}_s(k))\log (\frac{1}{2\pi}-\rho^{(j)}_s(k))\right]
}\,,
\label{eq:ratio}
\end{align}
where we explicitly reported the dependence of $r$ on $\nu$ and $\{a_i\}$. Using the constraint \eqref{eq:constraint} it is immediate to show that for $\nu=2$
\be
 \int_{0}^\pi\mathrm{d}k \sum_{j=1}^2\rho^{(j)}_s(k)\log \rho^{(j)}_s(k)= \frac{1}{2}\int_{0}^\pi\mathrm{d}k \sum_{j=1}^2\left\{\rho^{(j)}_s(k)\log \rho^{(j)}_s(k)+\left(\frac{1}{2\pi}-\rho^{(j)}_s(k)\right)\log \left(\frac{1}{2\pi}-\rho^{(j)}_s(k)\right)\right\}\,,
\ee
which implies 
\be
r(2;\{a_i\}) =\frac{1}{2}\,,\qquad\forall\, a_0,\,a_1\,.
\ee
For $\nu>2$, on the other hand, the ratio \eqref{eq:ratio} depends explicitly on $\{a_i\}$. This is demonstrated in Figs.~\ref{fig:entropy-ratio-numerics-nu-3}, \ref{fig:entropy-ratio-numerics-nu-4}, and \ref{fig:entropy-ratio-numerics-nu-5}, where we plot the different values of the ratio obtained by sampling many different configurations of $\{a_i\}$, for $\nu=3,4,$ and $5$ respectively. To perform the sampling, it is useful to explicitly solve the constraint in Eq.~\eqref{eq:FFinitial2} writing the moduli of the $\{a_i\}$ in polar coordinates and removing a redundant global phase. For example for $\nu=3$, we write the parameters as
\begin{align}
a_0&=\cos(\theta_1), &
a_1&=\sin(\theta_1)\cos(\theta_2) e^{\im \alpha_1}, &
a_2&=\sin(\theta_1)\sin(\theta_2) e^{\im \alpha_2}, &
\theta_i\in [0,\pi/2]&,\ \alpha_i\in[0,2\pi)\,.
\label{eq:parametrisation}
\end{align}
This allows us to simply sample the parameters at regular intervals in their allowed range.

\begin{figure}
\includegraphics[scale=0.8]{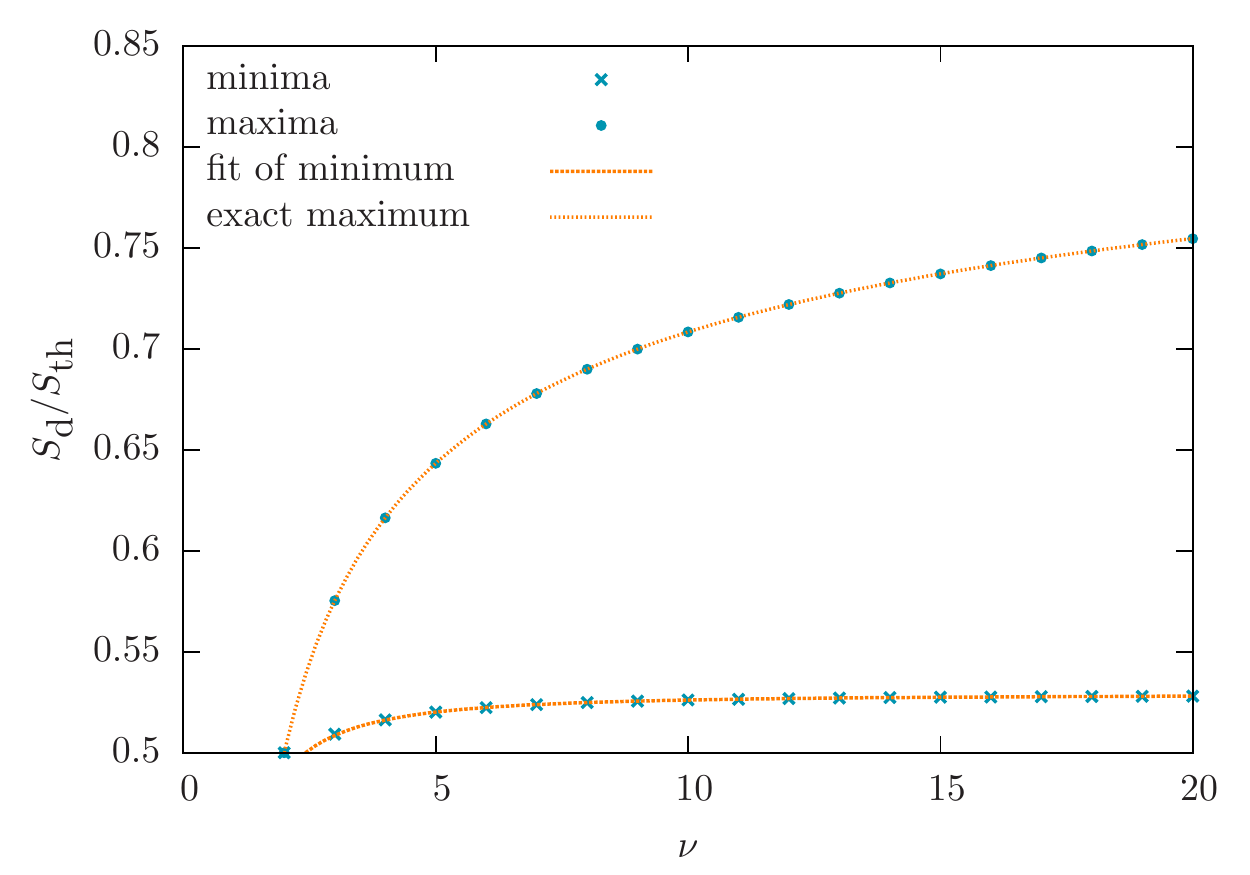}
\caption{This plot contains the extremal values of the ratio between the diagonal and thermodynamic entropies for $\nu=2,\ldots,20$. The dots are the maxima and minima obtained numerically using Mathematica. The upper curve is the exact maximum \eqref{eq:ratio-max}. The lower curve $r_{\rm min}(\nu) = 0.53-0.12\,\nu^{-1.6}$ was obtained using a fit in Mathematica on the data values at $\nu=10,\ldots,20$.  This graph demonstrates numerically that the ratio of these entropies lies between one half and one for $\nu\geq3$.}
\label{fig:entropy-ratio-extrema}
\end{figure}

From the plots we see that the values of $r(\nu;\{a_i\})$ are distributed between a maximum and a minimum,  $r_{\rm max}(\nu)$ and $r_{\rm min}(\nu)$, which are both larger than $1/2$ and smaller than $1$. To determine such maxima and minima analytically, one should extremise $r(\nu;\{a_i\})$ as a function of the parameters $\{a_i\}$ with the constraint in Eq.~\eqref{eq:FFinitial2}. This problem is not analytically tractable, but some analytical understanding can still be gained. It is convenient to consider the following functional 
\be
\Srat[\nu;\{\rho^{(j)}\}] = \frac{\Sred[\{\rho^{(j)}\}]}{\SYY[\{\rho^{(j)}\}]}\,,
\label{eq:ratiofunctional}
\ee  
together with the constraint 
\be
\sum_{j=1}^{\nu}\rho^{(j)}(k) = \frac{1}{2\pi}\,.
\label{eq:maximum}
\ee
In Appendix~\ref{app:extremisation} we show that the set of root densities 
\be
\rho^{(j)}(k)=\frac{1}{2\pi\nu}\,,\qquad\qquad\qquad j=1,\ldots,\nu\,,
\ee
is a local maximum for the functional. These functions correspond to $\{\rho^{(j)}_s(k)\}$ with 
\be
a_j=e^{i \phi}\qquad \text{and}\qquad a_i=0\qquad \forall\, i\neq j \qquad\text{for}\qquad j=0,\ldots,\nu-1\,,\qquad \phi\in \mathbb{R}\,,
\label{eq:amax}
\ee
\emph{cf}. Eq.~\eqref{eq:rhos}. This means that \eqref{eq:maximum} is also a local maximum for $r(\nu;\{a_i\})$. The value of the ratio at the point \eqref{eq:amax} reads as 
\be
r_*(\nu)\equiv r(\nu;\{\underbrace{0,\ldots,0,e^{i \phi}}_j,0,\ldots, 0\}) = \frac{\log\nu}{\nu\log\nu-(\nu-1)\log(\nu-1)}\,,\qquad \forall\,j=1,\ldots,\nu\,,\qquad \forall\,\phi\in \mathbb{R}\,.
\ee
Numerical maximisation in Mathematica of the function $r(\nu;\{a_i\})$ for $\nu=3,\ldots,20$ gives values coinciding with this formula, see Fig.~\ref{fig:entropy-ratio-extrema}. Furthermore, a numerical sweep across the possible initial states for $\nu=3,4,5$ (see Figures \ref{fig:entropy-ratio-numerics-nu-3}, \ref{fig:entropy-ratio-numerics-nu-4} and \ref{fig:entropy-ratio-numerics-nu-5}, respectively) indicates that it is in fact the global maximum. In other words, the numerical maximisation suggests 
\be
r_{\rm max}(\nu)=r_*(\nu)=\frac{\log\nu}{\nu\log\nu-(\nu-1)\log(\nu-1)}\,.
\label{eq:ratio-max}
\ee
Note that this global upper bound approaches one as $\nu$ increases
\be
\lim_{\nu\rightarrow\infty}\frac{\log\nu}{\nu\log\nu-(\nu-1)\log(\nu-1)}=1\,.  
\ee

To find the minimum, we turned to the numerical minimisation of Mathematica. The values of the minimum obtained are given in Table \ref{tab:ratio-min} and plotted in Figure \ref{fig:entropy-ratio-extrema}. The minimum for each $\nu>2$ is greater than one half, and it increases with increasing $\nu$. Our numerical results seem to suggest that the minimum approaches a constant $c\approx0.53$ in a power law fashion in the infinite $\nu$ limit (see Fig.~\ref{fig:entropy-ratio-extrema}). In Figures \ref{fig:entropy-ratio-numerics-nu-3}, \ref{fig:entropy-ratio-numerics-nu-4} and \ref{fig:entropy-ratio-numerics-nu-5} we verify that the ratio is never below the minimum value by sampling many different cases.

\begin{table}
\begin{tabular}{|c|c||c|c|}
\hline
$\nu$ & $r_{\rm min}$ & $\nu$ & $r_{\rm min}$\\
\hline
2 &  0.5000 & 12 & 0.5268\\
3 &  0.5092 & 13 & 0.5271\\ 
4 &  0.5163 & 14 & 0.5273\\
5 &  0.5201 & 15 & 0.5275\\
6 &  0.5224 & 16 & 0.5276\\
7 &  0.5238 & 17 & 0.5277\\
8 &  0.5248 & 18 & 0.5279\\
9 &  0.5256 & 19 & 0.5279\\
10 & 0.5261 & 20 & 0.5280\\ 
11 & 0.5265 & &\\
\hline
\end{tabular}
\caption{The values for the minimum value of the ratio between the diagonal and thermodynamic entropies for $\nu=2,\ldots,20$ obtained using the numerical minimisation function in Mathematica.}
\label{tab:ratio-min}
\end{table}

\section{Conclusions}
\label{sec:conclusions}

In this paper, we considered quantum quenches from a class of initial states which produce a distribution of elementary excitations with no pair structure, \emph{i.e.} no correlated pairs of particles with opposite momenta. Instead of pairs, such states create $\nu$-plets of elementary excitations with $\nu>2$. Our main goal has been to study the time evolution of the entanglement entropy and its stationary value. To limit the technical complications and maximise the analytical and numerical control we focused on a free fermionic model on the lattice. 

By comparing with high precision numerical results, only available for free models, we showed that the time evolution of the entropy after the quenches examined is not described by the well known semiclassical formula proposed in Ref.~\cite{calabrese_evolution_2005}. 
We showed that this is {\it not} due to a failure of the semiclassical interpretation but is caused by the breaking of the pair structure operated by our initial states. We proposed a generalised semiclassical theory which, in the scaling limit, exactly reproduces the numerical results. A key ingredient for the determination of such formula was the identification of the entanglement contributions carried by the different partitions of the correlated $\nu$-plets. As our generalised semiclassical description is expressed in a thermodynamic Bethe ansatz language, it is readily generalised to the interacting case in the spirit of Ref.~\cite{alba_entanglement_2016}. We expect this description to reproduce the dynamics of entanglement entropy after quenches from states generating distributions of correlated $\nu$-plets of quasi-particle excitations in generic integrable models, provided that such states exist.  

A neat prediction of our semiclassical theory is that in the infinite time limit the entanglement entropy of a region coincides with the thermodynamic entropy $S_{\rm th}$ of the local stationary state, in agreement with a currently widespread belief. We compared this value with that of the diagonal entropy $S_{\rm d}$~\cite{P:diagonalentropy} to check whether the relation $S_{\rm th}=2S_{\rm d}$, observed in many integrable models~\cite{leda-2014,kormos-2014, collura-2014, Gur14,F:finitesize, PVCR17,SPR:pre,D:escort}, is modified in the absence of the pair structure. We found that this is indeed the case. Specifically, we found that when the pair structure is not present the two quantities $S_{\rm d}$ and $S_{\rm th}$ are generically unrelated: their ratio depends on the details of the initial state, and for each $\nu>2$ its value ranges between a maximum and a minimum lying in the interval $(1/2,1]$. In thermodynamic Bethe ansatz language this result is ultimately due to the fact that the reduced entropy \eqref{eq:yangyangreduced} and the Yang-Yang entropy \eqref{eq:yang-yang-entropy} are two independent functionals for $\nu>2$. Since these functionals are not specific to the free case, we also expect this result to hold in the presence of integrable interactions.

\section{Acknowledgements}

We are grateful to Maurizio Fagotti for useful suggestions, valuable comments and for the helping us in the implementation of the numerical procedure used in Section~\ref{sec:entanglemententropy}. B. B. acknowledges the financial support by the ERC under the Advanced Grant 694544 OMNES and thanks SISSA for hospitality.

\appendix

\section{Wick's theorem for the states \eqref{eq:FFinitial2}}
\label{app:Wick}

Let us consider the $\nu\times\nu$ unitary matrix $\mathbb M$ such that 
\be
[\mathbb M]_{1j}= a_{\nu-j}\,,
\ee
where $\{a_{j}\}$ are the coefficients in \eqref{eq:FFinitial2}. Such matrix is constructed by putting as the rows the vectors of a basis of $\mathbb R^\nu$ which is orthonormal according to the canonical Hermitian scalar product and has as first element the vector $\underline v=(a_\nu,\ldots,a_0)$. We now use the matrix $\mathbb M$ as a building block to construct a block diagonal $L\times L$ unitary matrix $\mathbb U$ as follows
\be
\mathbb U=
\begin{pmatrix}
\mathbb M & 0 &  \dots  & 0 \\ 
0 & \mathbb M & \dots  & 0 \\
\vdots & \vdots & \ddots & 0\\
0 & 0  & \dots &  \mathbb M
\end{pmatrix}\,.
\ee 
The matrix is built of $(L/\nu)^2$ blocks (everything is well-defined because $L/\nu$ is an integer). 

Let us now use the matrix $\mathbb U$ to define a new set of fermions as follows 
\be
f^\dag_i = \sum_{j=1}^L [\mathbb U]_{ij} c^\dag_j\,.
\ee
Since $\mathbb U$ is unitary the fermions $\{f^\dag_i\}_{i=1}^{L}$ continue to satisfy the canonical anti-commutation relations \eqref{eq:CAR}. Finally, considering the mapping 
\begin{align}
e_i &= f^\dag_i\qquad i\in\{1,\nu+1,2\nu+1,\ldots,  \frac{L}{\nu} \nu-\nu+1\}\,,\\
e_i &= f_i \qquad i\in\{1,\ldots,L\}\setminus\{1,\nu+1,2\nu+1,\ldots, \frac{L}{\nu} \nu -\nu+1\}\,.
\end{align}
we find a set $\{e_i\}_{i=1}^L$ of fermions satisfying the canonical commutation relations \eqref{eq:CAR} such that 
\be
e_i \ket{\Phi^{\nu}_{\{a_1,\ldots,a_{\nu-1}\}}} =0 \qquad \forall\,\, i=1,...,L\,.
\ee
In other words, we proved that the fermionic operators satisfy Wick's theorem on the states $\ket{\Phi^{\nu}_{\{a_1,\ldots,a_{\nu-1}\}}}$~\cite{molinari_notes_2017}. 
\section{Restoration of one-site shift invariance}
\label{app:translational}

Since we are focusing on the time evolution of Gaussian initial states under a free Hamiltonian, the states remain Gaussian at all times. This means that the fermionic correlations can be decomposed using Wick's theorem and that to prove the restoration of one-site shift invariance it is sufficient to consider two-point functions
\begin{align}
\braket{\Phi^{\nu}_{\{a_m\}}|c^\dag_n(t)c_m(t)|\Phi^{\nu}_{\{a_m\}}}&=\frac{1}{L}\sum_{p,q}\braket{\Phi^{\nu}_{\{a_m\}}|\tilde c^\dag_p \tilde c^{\phantom{\dag}}_q|\Phi^{\nu}_{\{a_m\}}}e^{i(p n - q m)}e^{i(\varepsilon(p)  - \varepsilon(q))t}\,,\label{eq:corr1}\\
\braket{\Phi^{\nu}_{\{a_m\}}|c^\dag_n(t)c^\dag_m(t)|\Phi^{\nu}_{\{a_m\}}}&=\frac{1}{L}\sum_{p,q}\braket{\Phi^{\nu}_{\{a_m\}}|\tilde c^\dag_p \tilde c^{\dag}_q|\Phi^{\nu}_{\{a_m\}}}e^{i(p n + q m)}e^{i(\varepsilon(p)  + \varepsilon(q))t}\,.
\end{align}
The second function is identically zero in our case, since $\ket{{\Phi^{\nu}_{\{a_m\}}}}$ has a fixed number of particles, so it is trivially one-site shift invariant.  Focusing on the first one, we use the $\nu$-sites shift invariance to obtain
\be
\braket{\Phi^{\nu}_{\{a_m\}}|\tilde c^\dag_p \tilde c^{\phantom{\dag}}_q|\Phi^{\nu}_{\{a_m\}}}=\braket{\Phi^{\nu}_{\{a_m\}}|T^\nu\tilde c^\dag_p \tilde c^{\phantom{\dag}}_q(T^{\dag})^\nu|\Phi^{\nu}_{\{a_m\}}}=\braket{\Phi^{\nu}_{\{a_m\}}|\tilde c^\dag_p \tilde c^{\phantom{\dag}}_q|\Phi^{\nu}_{\{a_m\}}} e^{i (p-q)\nu}\,,
\ee
so that $\braket{\Phi^{\nu}_{\{a_m\}}|\tilde c^\dag_p \tilde c^{\phantom{\dag}}_q|\Phi^{\nu}_{\{a_m\}}}$ is nonzero only when 
\be
p= p_j(q)\equiv
\begin{cases}
\left(q+\frac{2\pi}{\nu}j\right)\!\!\!\!\!\mod 2 \pi \quad &\text{if}\quad \left(q+\frac{2\pi}{\nu}j\right)\!\!\!\!\!\mod 2 \pi \leq\pi\\
\left(q+\frac{2\pi}{\nu}j\right)\!\!\!\!\!\mod 2 \pi-2\pi \quad &\text{if}\quad \left(q+\frac{2\pi}{\nu}j\right)\!\!\!\!\!\mod 2 \pi \geq\pi
\end{cases}
\,,\qquad j=0,\ldots,\nu-1\,.
\ee
Plugging back in \eqref{eq:corr1} we have 
\begin{align}
\braket{\Phi^{\nu}_{\{a_m\}}|c^\dag_n(t)c_m(t)|\Phi^{\nu}_{\{a_m\}}}=&\frac{1}{L}\sum_{q}\sum_{j=0}^{\nu-1}\braket{\Phi^{\nu}_{\{a_m\}}|\tilde c^\dag_{p_j(q)} \tilde c^{\phantom{\dag}}_q|\Phi^{\nu}_{\{a_m\}}}e^{i(p_j(q)n - q m)}e^{i(\varepsilon(p_j(q))  - \varepsilon(q))t}\nonumber \\
=&\frac{1}{L}\sum_{q}\braket{\Phi^{\nu}_{\{a_m\}}|\tilde c^\dag_{q} \tilde c^{\phantom{\dag}}_q|\Phi^{\nu}_{\{a_m\}}}e^{i q(n - m)} \nonumber \\
&+\frac{1}{L}\sum_{q}\sum_{j=1}^{\nu-1}\braket{\Phi^{\nu}_{\{a_m\}}|\tilde c^\dag_{p_j(q)} \tilde c^{\phantom{\dag}}_q|\Phi^{\nu}_{\{a_m\}}}e^{i((q+\frac{2\pi}{\nu}j)n - q m)}e^{i(\varepsilon(q+\frac{2\pi}{\nu}j)  - \varepsilon(q))t}\,.\label{eq:negligible}
\end{align}
Taking first the thermodynamic limit and then the infinite time limit we then find 
\begin{align}
\lim_{t\rightarrow\infty}{\textstyle\lim_{\rm th}}\braket{\Phi^{\nu}_{\{a_m\}}|c^\dag_n(t)c_m(t)|\Phi^{\nu}_{\{a_m\}}}=\int_{-\pi}^\pi\frac{{\rm d}q}{2\pi}\braket{\Phi^{\nu}_{\{a_m\}}|\tilde c^\dag_{q} \tilde c^{\phantom{\dag}}_q|\Phi^{\nu}_{\{a_m\}}}e^{i q(n - m)}\,,
\label{eq:translinv}
\end{align}
where we used the saddle point approximation noting that 
\be
\varepsilon\left(q+\frac{2\pi}{\nu}j\right)  - \varepsilon(q)=-2J\cos(q)\left(\cos\left(\frac{2\pi}{\nu}j\right)-1\right)+2 J\sin(q) \sin\left(\frac{2\pi}{\nu}j\right)\,,
\ee
which is not a constant function of $q$ for each $j=1,\ldots,\nu-1$. This means that the thermodynamic limit of the second term on the right-hand side of \eqref{eq:negligible} contributes at most as $t^{-1/2}$ for large times. From Eq.~\eqref{eq:translinv}, we see that the fermionic two-point function in the infinite time limit becomes invariant under one-site shifts. This concludes the proof.

\section{Extremisation of \eqref{eq:ratiofunctional}}
\label{app:extremisation}

Here we extremise the functional $\Srat[\nu;\{\rho^{(j)}\}]$ with the constraint \eqref{eq:constraint}. The constraint is immediately solved by writing $\rho^{(\nu)}(k)$ in terms of the other root densities. We begin by taking the variation of $\Srat[\nu;\{\rho^{(j)}\}]$ with respect to $\rho^{(i)}(k)$ for each $i\in\{1,2,\ldots,\nu-1\}$
\begin{align}
\frac{\delta \Srat[\nu;\{\rho^{(j)}\}]}{\delta\rho^{(i)}(k)} &= 
\frac{\delta\Sred[\{\rho^{(j)}\}]}{\delta\rho^{(i)}(k)}\frac{1}{\SYY[\{\rho^{(j)}\}]}
- \frac{\delta\SYY[\{\rho^{(j)}\}]}{\delta\rho^{(i)}(k)}\frac{\Sred[\{\rho^{(j)}\}]}{\SYY^2[\{\rho^{(j)}\}]}\,.
\end{align}
The variations of the  reduced and thermodynamic entropies are
\begin{align}
\frac{\delta\Sred[\{\rho^{(j)}\}]}{\delta\rho^{(i)}(k)} &=
\log\left(\frac{\rho^{(\nu)}(k)}{\rho^{(i)}(k)} \right)
&
\frac{\delta\SYY[\{\rho^{(j)}\}]}{\delta\rho^{(i)}(k)} &=
\log\left(\frac{\rho^{(\nu)}(k)}{\rho^{(i)}(k)} \right)
+ \log\left(\frac{\frac{1}{2\pi}-\rho^{(i)}(k)}{\frac{1}{2\pi}-\rho^{(\nu)}(k)} \right)\,.
\end{align}
Setting the variation of $\Srat$ to zero gives the following relation
\begin{align}
\log\left(\frac{\rho^{(\nu)}_*(k)}{\rho^{(i)}_*(k)} \right)\frac{1}{\SYY[\{\rho^{(j)}\}]}
-\left[\log\left(\frac{\rho^{(\nu)}_*(k)}{\rho^{(i)}_*(k)} \right)
+ \log\left(\frac{\frac{1}{2\pi}-\rho^{(i)}_*(k)}{\frac{1}{2\pi}-\rho^{(\nu)}_*(k)} \right) \right]\frac{r_*}{\SYY[\{\rho^{(j)}\}]}=0\,,
\end{align}
where we let $r_*=\Srat[\nu;\{\rho^{(j)}_*\}]$. This expression simplifies to the following condition for the extremand root densities
\be
\left[\rho^{(i)}_*(k)\right]^{1-r_*} \left[\frac{1}{2\pi}-\rho^{(i)}_*(k)\right]^{r_*}
= \left[\rho^{(\nu)}_*(k)\right]^{1-r_*} \left[\frac{1}{2\pi}-\rho^{(\nu)}_*(k)\right]^{r_*}, \qquad
i=1,\ldots,\nu-1\,.
\ee
A simple solution to this relation is when all the root densities are equal and constant
\be
\rho^{(1)}_*(k) = \ldots = \rho^{(\nu)}_*(k) = c, \qquad c\in \mathbb{R}\,.
\ee
The constraint \eqref{eq:constraint} fixes the value of this constant to be
\be
\rho^{(1)}_*(k) = \ldots = \rho^{(\nu)}_*(k) = \frac{1}{2\pi\nu}\,.
\label{eq:solution}
\ee
This state is of the form of the steady state \eqref{eq:rhos}, where all the $A_n=0$, which can be obtained by taking 
\be
a_j=e^{i \phi}\qquad \text{and}\qquad a_i=0\qquad \forall\, i\neq j \qquad\text{for}\qquad j=1,\ldots,\nu\,,\qquad \phi\in \mathbb{R}\,,
\ee
in the initial state \eqref{eq:FFinitial2}. Therefore, it corresponds to a local extremum of the ratio of the diagonal and thermodynamic entropies in the steady state \eqref{eq:rhos}. The value of the ratio at this local extremum is
\be
r_* = \frac{\log\nu}{\nu\log\nu-(\nu-1)\log(\nu-1)}\,.
\label{eq:ratio-maxapp}
\ee
To determine the type of local extremum, we compute the second variation around the solution \eqref{eq:solution}. It reads as
\be
\delta^2\Srat[\nu;\delta\rho^{(1)}, \ldots, \delta\rho^{(\nu-1)}]= -\frac{\frac{2\pi\nu}{\nu-1}((r_*+1)\nu-1)}{\log\frac{\nu}{\nu-1}+\frac{1}{\nu}\log(\nu-1)}\int_{\pi-\frac{2\pi}{\nu}}^{\pi} {\rm d} k \sum_{i,j=1}^{\nu-1} (\delta_{i,j}+1)\delta\rho^{(i)}(k)\delta\rho^{(j)}(k)\,.
\label{eq:secondvariation}
\ee
The the matrix $M$ with elements $M_{i,j}=\delta_{i,j}+1$ is positive definite, as it can be easily proven computing the eigenvalues
\be
\lambda_1=\nu\,, \qquad \qquad\lambda_j = 1\,,\qquad\qquad j=2,\ldots,\nu-1\,.
\ee
Since 
\be
-\frac{\frac{2\pi\nu}{\nu-1}((r_*+1)\nu-1)}{\log\frac{\nu}{\nu-1}+\frac{1}{\nu}\log(\nu-1)}<0\,,
\ee
the second variation \eqref{eq:secondvariation} is strongly negative (\emph{i.e.} its negative is strongly positive according to the definition in Theorem 2 on page 100 of~\cite{calculusofvariations})
\be
\delta^2\Srat[\nu;\delta\rho^{(1)}, \ldots, \delta\rho^{(\nu-1)}]<- A \sum_{i=1}^{\nu-1}\int_{\pi-\frac{2\pi}{\nu}}^{\pi} {\rm d} k (\delta\rho^{(i)}(k))^2\qquad\qquad A>0\,.
\ee
This implies that the local extremum \eqref{eq:ratio-maxapp} is a local maximum~\cite{calculusofvariations}.


\begin{thebibliography}{99}
\bibitem{note}
This particular ordering can never be realised in the XX model, but we nonetheless consider it  for illustrative purposes.
\end{thebibliography}
\end{document}